\documentclass{CUP-JNL-DTM}%
\usepackage{xr}
\makeatletter
\newcommand*{\addFileDependency}[1]{
\typeout{(#1)}
\@addtofilelist{#1}
\IfFileExists{#1}{}{\typeout{No file #1.}}
}\makeatother

\newcommand*{\myexternaldocument}[1]{%
\externaldocument{#1}%
\addFileDependency{#1.tex}%
\addFileDependency{#1.aux}%
}
\myexternaldocument{supplementary}

\usepackage{graphicx}
\usepackage{multicol,multirow}
\usepackage{amsmath,amssymb,amsfonts}
\usepackage{mathrsfs}
\usepackage{amsthm}
\usepackage{rotating}
\usepackage{appendix}
\usepackage{subfiles}
\usepackage[natbib,style=apa]{biblatex}
\usepackage{tabularx}
\usepackage{ifpdf}
\usepackage[T1]{fontenc}
\usepackage{newtxtext}
\usepackage{newtxmath}
\usepackage{textcomp}
\usepackage{xcolor}
\usepackage{lipsum}
\usepackage{xr}
\usepackage[colorlinks,allcolors=blue]{hyperref}

\usepackage{wrapfig}
\numberwithin{equation}{section}

\usepackage{url} 
\usepackage{lineno}
\usepackage{tabularray}
\usepackage{amsmath}
\usepackage{algorithm2e}
\usepackage{graphicx}
\usepackage{bm}
\usepackage{float}

\usepackage{soul}

\newcommand{\mycite}[1]{(\cite{#1})}
\newcommand{\cgfac}{\kappa}
\newcommand{\cotwo}{$\text{CO}_2$}

\newcommand{\hres}[1]{#1^{\uparrow}}
\newcommand{\inv}[1]{#1_{\operatorname{inv}}}

\newcommand{\filter}[1]{\overline{#1}}

\newcommand{\cgvar}[1]{\big\langle #1\big\rangle_{\cgfac}}

\newcommand{\cnnglobal}{CNN (global)}
\newcommand{\cnnfoureg}{CNN (4 regions)}

\newcommand{\rsquare}[1]{$\text{R}^2$}

\jname{Enviromental Data Science}
\artid{20}
\jyear{2023}
\jvol{4}
\jissue{1}
\usepackage{subfiles}

\begin{document}

\begin{Frontmatter}

\title[Article Title]{An Analysis of Deep Learning Parameterizations for Ocean Subgrid Eddy Forcing}


\author[1]{Cem Gultekin}
\author[1]{Adam Subel}
\author[3]{Cheng Zhang}
\author[1]{Matan Leibovich}
\author[1]{Pavel Perezhogin}
\author[3]{Alistair Adcroft}
\author[1,2]{Carlos Fernandez-Granda}
\author[1]{Laure Zanna}

\address[1]{Courant Institute of Mathematical Sciences, New York University, New York, NY 10012, USA}
\address[2]{Center for Data Science, New York University, New York, NY 10011, USA}
\address[3]{Program in Atmospheric and Oceanic Sciences, Princeton University, Princeton, NJ 08542, USA}

\authormark{Cem Gultekin et al.}



\abstract{
Due to computational constraints, climate simulations cannot resolve a range of small-scale physical processes, which have a significant impact on the large-scale evolution of the climate system. Parameterization is an approach to capture the effect of these processes, without resolving them explicitly. In recent years, data-driven parameterizations based on convolutional neural networks have obtained promising results. In this work, we provide an in-depth analysis of these parameterizations developed using data from ocean simulations. The parametrizations account for the effect of mesoscale eddies toward improving simulations of momentum, heat, and mass exchange in the ocean. Our results provide several insights into the properties of data-driven parameterizations based on neural networks. First, their performance can be substantially improved by increasing the geographic extent of the training data. Second, they learn nonlinear structure, since they are able to outperform a linear baseline. Third, they generalize robustly across different \cotwo{} forcings, but not necessarily across different ocean depths. Fourth, they exploit a relatively small region of their input to generate their output. Our results will guide the further development of ocean mesoscale eddy parameterizations, and multiscale modeling more generally.
}
\end{Frontmatter}

\section*{Impact Statement}
Climate simulations spanning decades or centuries cannot be run at a high spatial resolution due to computational constraints. Unfortunately, as a result, critical fine-scale physical processes such as mesoscale eddies are not captured by the simulations. Deep learning has emerged as a promising solution to incorporate the effect of these processes. In this work, we evaluate several properties of these models that are crucial for their deployment in climate simulations.

\section{Introduction}\label{sec:introduction}

Despite advances in hardware, climate simulations spanning decades or centuries are limited in their spatial resolution to tens of kilometers \mycite{balaji2021climbing}. This resolution is insufficient to resolve mesoscale eddies that are crucial for the exchange of momentum, heat, and mass \mycite{capet2008mesoscale,salmon1980baroclinic} as well as for large-scale ocean circulation~\mycite{hallberg2013using, keating2012new, waterman2011eddy}. 
Parameterization is an approach to capture the effect of small-scale processes on the large-scale variables in climate models without resolving them explicitly. The main idea is to modify the governing equations of a low-resolution model by adding a forcing term called subgrid forcing. This term encapsulates the \emph{missing physics}, representing the effect of the unresolved physical processes on the resolved variables in the climate model. The key challenge is that the subgrid forcing is a function of high-resolution quantities, but only low-resolution quantities are observed in the model. Traditional parameterization approaches are typically rooted in first-principle analysis of the climate-model physics \mycite{randall2003breaking, bony2015clouds, schneider2017climate}.

In recent years, data-driven parameterizations have been developed with promising results~\mycite{yuval2020stable,beucler2021enforcing,frezat2021physical,GUAN2022111090, zanna2020data, guillaumin2021stochastic,zhang2023implementation,sane2023verticalmixing,bodner2023submeso, perezhogin2023implementation}. In the data-driven framework, a high-resolution simulation capable of resolving the physical processes of interest is utilized as ground truth. The resolution of the data is then artificially reduced via filtering and coarse-graining to enable the computation of "ground-truth" subgrid forcing. Machine learning (ML) algorithms are used to predict this forcing from the low-resolution data. 
In the case of mesoscale eddies, applying this framework to convolutional neural networks (CNNs) has achieved promising results \mycite{zanna2020data,guillaumin2021stochastic}. 

Our goal in this work is to provide an in-depth study of CNNs for parameterization. We utilize the CM2.6 dataset as our source of high-resolution data. CM2.6 is a publicly-available advanced coupled climate model \mycite{griffies2015handbook} with an approximate resolution of 0.1$^{\circ}$, making it well-suited for accurately representing mesoscale eddies \mycite{HALLBERG201392}. The dataset includes surface and subsurface data for two levels of \cotwo{} in the atmosphere.  
\cite{guillaumin2021stochastic} proposed a CNN-parameterization for surface momentum fields in CM2.6 at $0.4^{\circ}$ resolution. 
Here, we extend their approach to parameterize temperature, and
build upon it to study the following key questions about CNN-based mesoscale-eddy parameterizations. 

 \noindent \textbf{Does the geographic extent of the training dataset matter?} In  \citet{guillaumin2021stochastic}, the training domain was constrained to four relatively small regions of the Pacific and Atlantic Oceans. We show that extending the training dataset to encompass the entire global ocean surface leads to a substantial improvement in performance, particularly for temperature.

\noindent \textbf{Do the properties of CNN parameterizations change at different resolutions?} When training a CNN parameterization, an important consideration is how to choose the resolution of the target low-resolution climate model. We studied the performance of CNN-based parameterizations at resolutions of $0.4^{\circ}$, $0.8^{\circ}$, $1.2^{\circ}$, and $1.6^{\circ}$. Our results show to what extent reducing resolution decreases the skill of the parameterizations, and also demonstrate that their properties remain similar across the different resolutions.

\noindent \textbf{Do CNN-parameterizations just \emph{invert} the coarse-graining and filtering operator?} As explained above, CNN-parameterizations estimate subgrid forcing from low-resolution data, obtained from a high-resolution model via filtering and coarse-graining. These filtering and coarse-graining operators are linear and can be partially inverted to estimate subgrid turbulence fields (and hence the subgrid forcing itself)  \mycite{langford1999optimal}. Examples of subgrid parameterizations based on partial filter inversion include the velocity gradient model 
\mycite{clark1979evaluation, chow2005explicit}, the scale-similarity model \mycite{bardina1980improved, meneveau2000scale} and the approximate deconvolution model \mycite{stolz1999approximate}. 
An important question is whether CNN-parameterizations just learn to implement this partial inversion. We developed a baseline parameterization solely based on linear inversion to answer it. Our results indicate that CNNs outperform this baseline parameterization, suggesting that they are able to leverage physical structure and do not merely learn to perform inversion. 

\noindent \textbf{Do data-driven parameterizations generalize across \cotwo{} levels and ocean depths?} Understanding the generalization properties of data-driven parameterizations is crucial for their deployment in realistic models. In particular, determining their behavior at different \cotwo{} levels is critical for long-term climate simulations and predictions. Our results show that a CNN-based parameterization trained at pre-industrial \cotwo{} levels generalizes robustly at significantly increased \cotwo{} levels.  In addition, we investigate the generalization ability these models across different ocean depths, ranging from the surface down to 728 meters. In MOM-suite climate models, the ocean is represented as vertically stacked, irregularly spaced horizontal layers \mycite{griffies2015handbook}, with substantially different dynamics. At the surface, wind, solar radiation, and atmospheric interactions dominate circulation, whereas at greater depths, dynamics depend more on temperature and salinity \mycite{salmon1980baroclinic}. Our results show that surface-trained models have difficulties generalizing at greater depths, and, conversely, models trained beyond 55 meters do not generalize robustly to the surface.

\noindent \textbf{What spatial extent do CNN-parameterizations require as input?} The spatial extent needed to compute a parameterization at each grid point greatly influences its computational cost. In addition, it informs the degree of nonlocality needed to approximate the local subgrid forcing. In order to investigate its impact on CNN-parameterizations, we evaluated the performance of multiple CNNs with a range of input sizes. These experiments were complemented with a gradient-based sensitivity analysis. Our results indicate that a much smaller input size than the one used in previous models is sufficient to achieve strong performance.

The paper is organized into five sections, including the introduction. In Section \ref{sec:data-generation}, we describe the governing equations of the climate simulation, and explain how we generate the data to train the data-driven parameterizations. In Section~\ref{sec:methodology}, we define data-driven parameterizations based on deep learning, as well as a baseline parameterization based on linear inversion. In Section \ref{sec:experiments}, we describe our experiments and present the results. Finally, in Section \ref{sec:discussion-outlook}, we summarize our conclusions and discuss directions for future research.

 \section{Governing Equations and Data Coarsening }\label{sec:data-generation}
In this section we describe the governing equations of our climate model of interest, as well as the data-coarsening procedure used to generate the data needed to train data-driven parameterizations.  
Our \emph{ground-truth} high-resolution climate model is CM2.6, a coupled model that includes various physical quantities describing oceanic evolution, such as momentum, temperature, salinity, and biogeochemical tracers \mycite{griffies2015handbook}. The model resolution ($0.1^{\circ}$) is sufficient to faithfully represent mesoscale eddies. In our experiments, we consider seven different ocean depths, including the surface, and two different \cotwo{} levels. 

Our study focuses on the evolution of three variables: longitudinal and latitudinal momentum fields ($u$ and $v$, respectively) and the temperature field ($T$). The evolution of these quantities is governed by the following equation:

\begin{equation}\label{eq:euler_2d}
     \frac{\partial {\hres{c}}}{\partial t}
    +{\hres{\bm{u}}}\cdot \nabla {\hres{c}} = \hres{F_{c}},\quad c\in \{u,v,T\}
\end{equation}

We use the arrow symbol $\hres{}$ to denote fine-grid variables. The vector $\hres{\bm{u}}$ represents the components of longitudinal ($u$) and latitudinal ($v$) momentum on the fine grid. 
The gradient operator $\nabla$ contains the spatial derivatives in these two directions using spherical coordinates. 
Note that we are omitting the discussion of the vertical momentum contribution to the equation, merely because its contribution to the subgrid terms diagnosed in the model is small compared to the horizontal terms. 
The momentum forcing terms $\hres{F_{u}}$ and $\hres{F_{v}}$ account for external sources of dynamics, such as frictional and pressure stresses, the Coriolis force, or transport between vertically adjacent stratified ocean layers \mycite{griffies2015handbook}. The temperature forcing $\hres{F_T}$ accounts for isopycnal and diapycnal mixing, as well as heat fluxes at the ocean surface.

In climate modeling, it is common to employ a local Cartesian coordinate system that approximates the Earth's surface as a flat plane when dealing with small variations. This coordinate system allows for measurements in meters rather than angles. 
Therefore, we define $\nabla = (\partial_x,\partial_y)$ as the spatial gradient in meters along the longitudinal and latitudinal directions. 

Our goal is to study the properties of data-driven parameterizations that estimate the subgrid forcing of low-resolution versions of CM2.6. To obtain these coarser datasets, we reduce the resolution of the CM2.6 data  via a two-step procedure proposed in \citet{guillaumin2021stochastic}, consisting of filtering and coarse-graining. Filtering smooths the data on the original high-resolution fine grid. We leverage two alternative filters for this purpose: Gaussian filtering and General-Circulation-Model (GCM) filtering~\mycite{Loose2022}. 

We use a tilde  $\filter{\phantom{a}}$ to denote the low-resolution variables. The equations governing the low-resolution variables are derived from \eqref{eq:euler_2d} by applying filtering on both sides. 

The vertical advection term ($w\frac{\partial c}{\partial z}$) is significantly smaller than the horizontal terms, at all depths in this dataset. Therefore we reduce our analysis to the 2D system: 

In order to achieve the same form as the high-resolution equation, the low-resolution advective term, $\filter{\hres{\bm{u}}}\cdot \nabla \filter{\hres{c}}$ is added to both sides and the terms are rearranged: 

\begin{subequations}
    \begin{align}
        \filter{\big(\frac{\partial 
        \hres{c}
        }{\partial t} + \hres{\bm{u}}\cdot \nabla \hres{c}\big)} &= \filter{\hres{F_c}},\\
        \frac{\partial \filter{\hres{c}}}{\partial t}
        + \filter{\hres{\bm{u}}}\cdot \nabla\filter{\hres{c}}
        & = \filter{\hres{\bm{u}}}\cdot \nabla\filter{\hres{c}} - \filter{\big(\hres{\bm{u}}\cdot \nabla \hres{c}\big)}+
        \filter{\hres{F_c}}\label{eq:euler-low-res-before-coarse-graining}\\
        \hres{S_c} &\equiv 
        \filter{\hres{\bm{u}}}\cdot \nabla\filter{\hres{c}} - \filter{\big(\hres{\bm{u}}\cdot \nabla \hres{c}\big)},\quad c\in \{u,v,T\}
    \end{align}
\end{subequations}

\begin{figure}[!h]
\begin{tabularx}{0.95\textwidth} { 
>{\centering\arraybackslash}m{0.005\linewidth} 
>{\centering\arraybackslash}m{0.2\linewidth}
>{\centering\arraybackslash}m{0.2\linewidth}
>{\centering\arraybackslash}m{0.2\linewidth} 
>{\centering\arraybackslash}m{0.2\linewidth}
>{\centering\arraybackslash}m{0.025\linewidth}}
  & $\quad\cgfac = 4$ 
  & $\quad\cgfac = 8$ 
  & $\quad\quad\cgfac = 12$ 
   &$\quad\quad\cgfac = 16$ 
  &  \\
 ${u}$ & 
\includegraphics[width=1.1\linewidth]{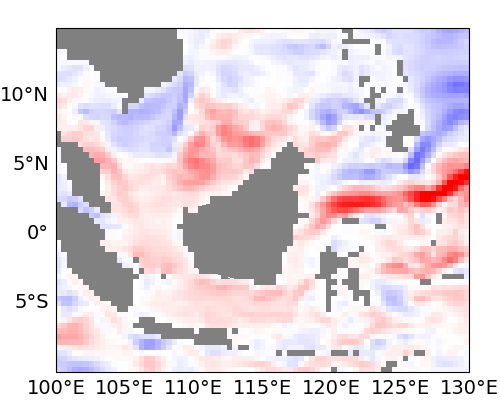}& 
 \includegraphics[width=1.1\linewidth]{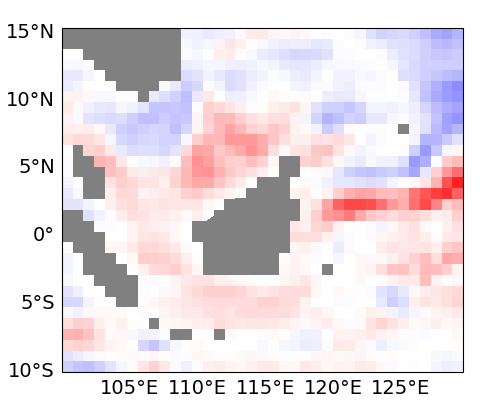}& 
 \includegraphics[width=1.1\linewidth]{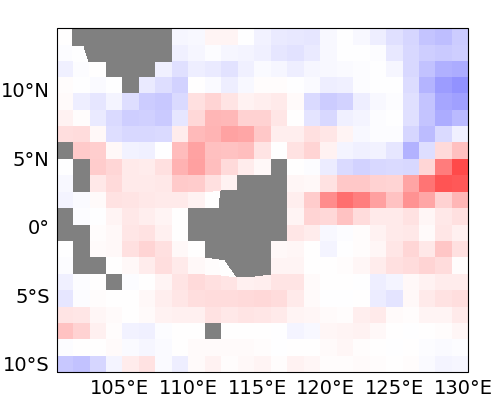}&
 \includegraphics[width=1.1\linewidth]{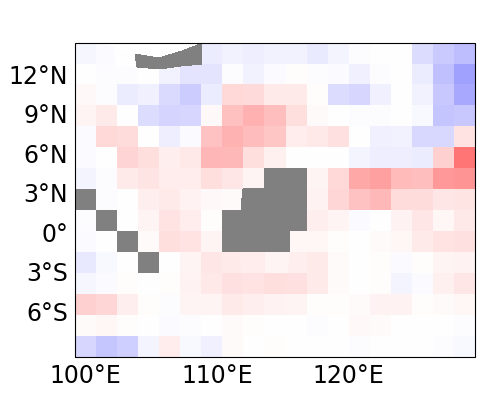}&
 \includegraphics[width=3\linewidth]{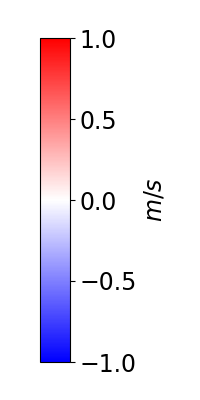}\\
 ${T}$ &
 \includegraphics[width=1.1\linewidth]{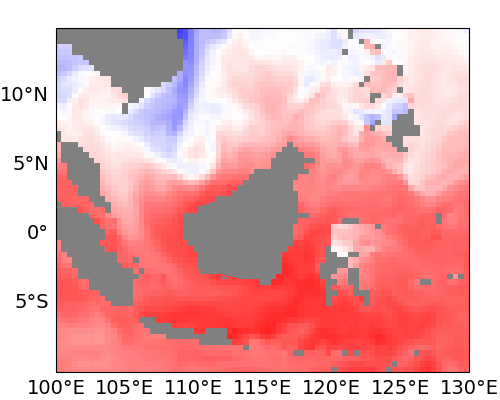}& 
 \includegraphics[width=1.1\linewidth]{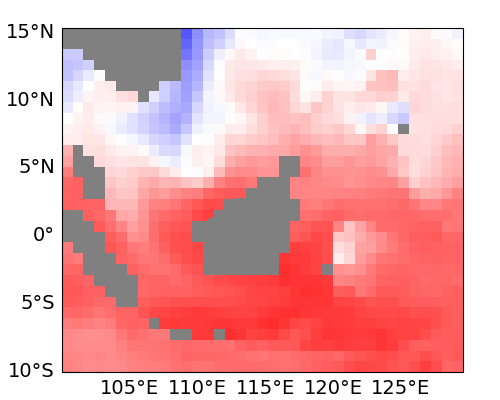}& 
 \includegraphics[width=1.1\linewidth]{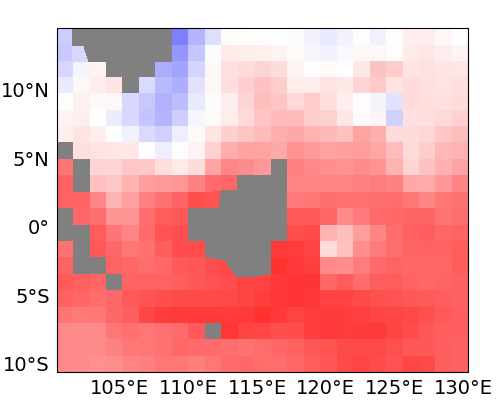}&
 \includegraphics[width=1.1\linewidth]{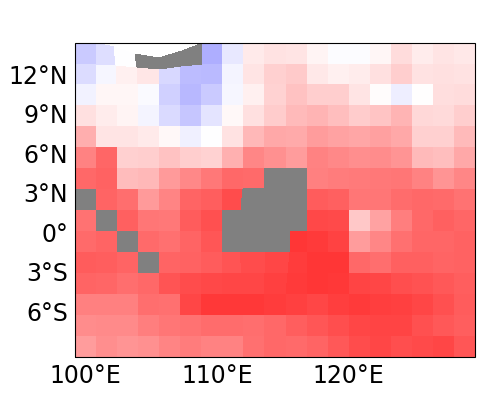}&
 \includegraphics[width=3\linewidth]{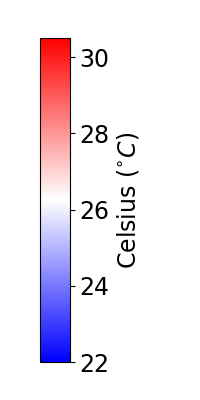}\\
 $S_u$ &
 \includegraphics[width=1.1\linewidth]{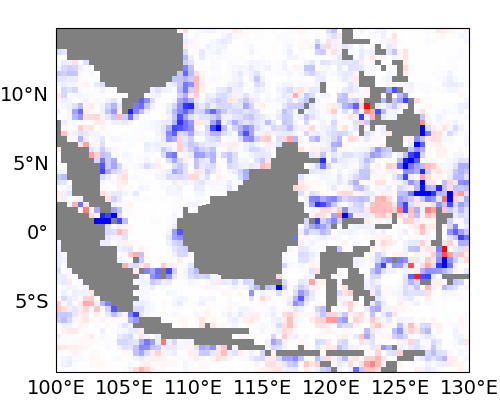}& 
 \includegraphics[width=1.1\linewidth]{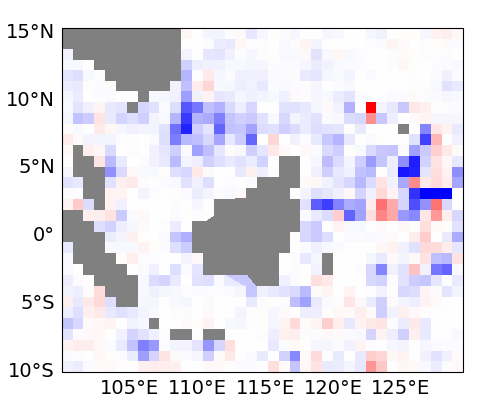}& 
 \includegraphics[width=1.1\linewidth]{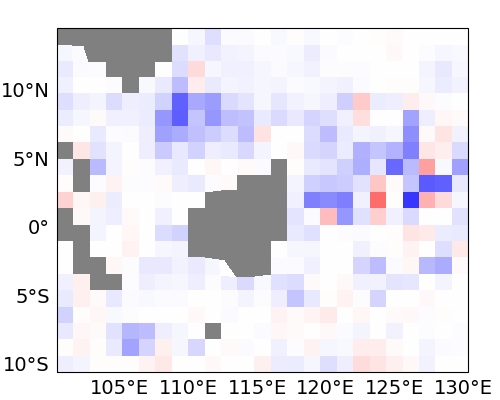}&
 \includegraphics[width=1.1\linewidth]{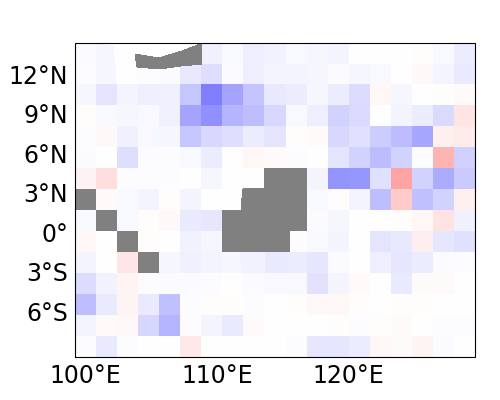}&
 \includegraphics[width=3\linewidth]{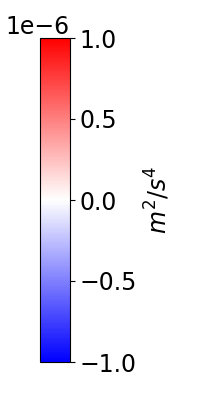}\\
 $S_T$ &
 \includegraphics[width=1.1\linewidth]{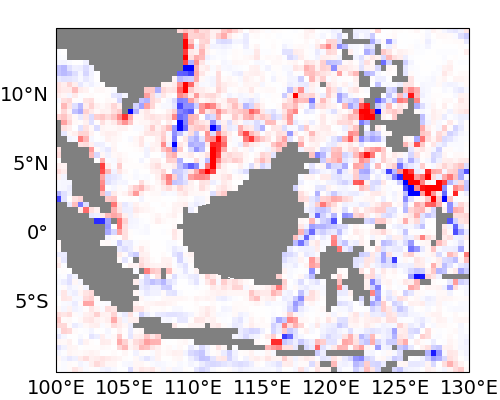}& 
 \includegraphics[width=1.1\linewidth]{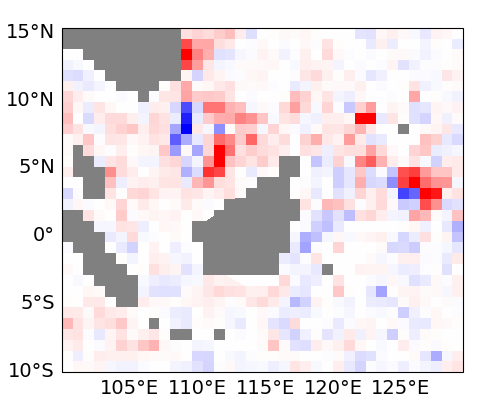}& 
 \includegraphics[width=1.1\linewidth]{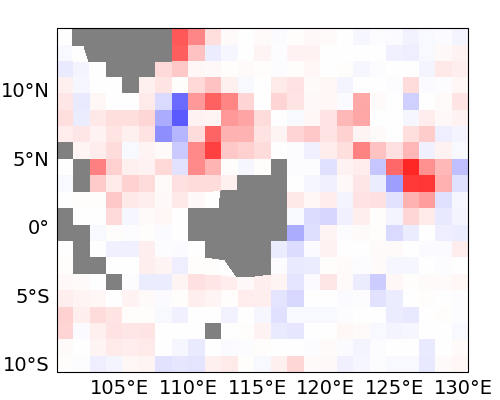}&
 \includegraphics[width=1.1\linewidth]{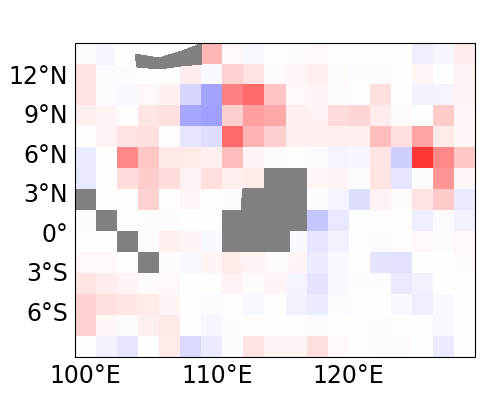}&
 \includegraphics[width=3\linewidth]{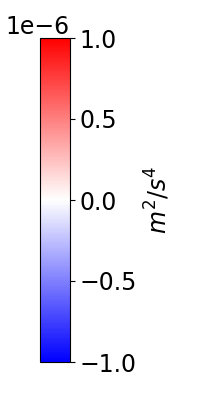}\\
\end{tabularx}
\caption{Examples of coarse-grid variables for different values of the coarsening factor $\cgfac$ along with their corresponding subgrid eddy forcing. The fine-grid data are extracted from the CM2.6 surface dataset, and filtered using GCM filtering. The rows represent longitudinal momentum (${u}$), temperature (${T}$), subgrid longitudinal momentum forcing ($S_u$), and subgrid temperature forcing ($S_T$) respectively} 

    \label{fig:subgrid-eddy-forcing-snapshot}
\end{figure}

The term $\hres{S_c}$ is the \emph{subgrid forcing} on the fine-grid. The low resolution data are mapped to a coarser grid via coarse-graining which reduces the grid dimensions by a coarse-graining factor $\cgfac$. This yields the coarse-grid dataset. We use $\cgvar{}$ to denote the coarse-graining operation. For example, the longitudinal momentum on the coarse grid equals 
$$u = \cgvar{\filter{\hres{u}}}.$$ 
Crucially, the coarse-grained subgrid forcing $S_{c}$, 
\begin{align}
    S_{c} = \cgvar{\filter{\hres{\bm{u}}}\cdot \nabla \filter{\hres{c}} - \filter{\hres{\bm{u}} \cdot \nabla \hres{c}}}, &\quad c = \{u,v,T\},     \label{eq:subgrid-forcing-definition}
    \end{align}
    depicted in Figure~\ref{fig:subgrid-eddy-forcing-snapshot}, depends on the fine-grained quantities. The key challenge of parameterization is to estimate this subgrid forcing from coarse-grained data, in order to \emph{close} the corresponding coarse-grid equations. 

 \section{Methodology}\label{sec:methodology}
 
 \subsection{Data-Driven Parameterization Based on Deep Learning}\label{sec:data-driven}

In this section, we describe a data-driven framework for the estimation of the subgrid forcing defined in Section~\ref{sec:data-generation} using deep neural networks. To simplify the exposition, let $S_{t,x}$ indicate the subgrid forcing corresponding to one of our three variables of interest (longitudinal momentum, latitudinal momentum, or temperature) at a time $t$ and location $x$. The goal is to estimate $S_{t,x}$ from the low-resolution momentum and temperature fields, denoted by $\mathbf{u}_{t,x}$ and $T_{t,x}$, surrounding that location at that time. 

Following previous studies \mycite{bolton2019applications,ross2023benchmarking,guillaumin2021stochastic, perezhogin2023generative}, we used a deep convolutional neural network $\widehat{S}_{\Theta}$, where $\Theta$ represents the network parameters, to produce an estimate $\widehat{S}_{\Theta}(\mathbf{u}_{t,x},T_{t,x})$ of the subgrid forcing $S_{t,x}$ given the low-resolution data $\mathbf{u}_{t,x}$ and $T_{t,x}$. In order to train and evaluate the neural network, we partitioned the CM2.6 data in time: the first 80\% was used to create a training set $\mathcal{T}_{\operatorname{train}}$ and the last 15\% to create a test set $\mathcal{T}_{\operatorname{test}}$, leaving a 5\% buffer in between to minimize dependence between the sets due to autocorrelation. 
The training and test examples were generated from the high-resolution CM2.6 data: $\mathbf{u}_{t,x}$ and $T_{t,x}$ were obtained via filtering and coarse-graining of the high-resolution momentum and temperature maps, and the corresponding ground-truth subgrid forcing $S_{t,x}$ was computed following \eqref{eq:subgrid-forcing-definition}. These data correspond to a control simulation with pre-industrial atmospheric \cotwo{} levels. 

In addition, a forced simulation with increased \cotwo{} levels was used to create another test set. The forced simulation experiences a one percent increase in \cotwo{} levels per year from the levels of the control simulation until it reaches double \cotwo{} levels after 70 years. The available data from the simulation corresponds to years 60–80. These data were processed in the same way as data from the control simulation to create the $+1\%$\cotwo{} test dataset.

Additional training and test sets were created from subsurface data at different ocean depths. These data from the simulation at control \cotwo{} levels, are available in the form of 5-day averages over the same time period. They were processed in the same way as the surface data to generate corresponding training and test sets for each depth. 

The parameters of the neural network were learned via minimization of a loss function, quantifying the approximation error over the training set. We consider two different loss functions. The first loss is the mean squared error (MSE), approximated by the residual sum of squared errors,  
\begin{align}
\text{MSE}(\Theta) = \frac{1}{n_{\operatorname{train}}} \sum_{t \in \mathcal{T}_{\operatorname{train}}} \sum_{x 
\in \mathcal{X}} \left( \widehat{S}_{\Theta}(\mathbf{u}_{t,x},T_{t,x}) - S_{t,x} \right)^2, 
\end{align}
where $\mathcal{T}_{\operatorname{train}}$, $\mathcal{X}$ and $n_{\operatorname{train}}$ denote the times, locations and number of data included in the training set. The second loss is the heteroscedastic Gaussian loss (HGL) proposed in \cite{guillaumin2021stochastic}, which simultaneously estimates the conditional mean and conditional variance of the subgrid forcing by maximizing a Gaussian likelihood
\begin{align}
\text{HGL}(\Theta) =  \sum_{t \in \mathcal{T}_{\operatorname{train}}} \sum_{x 
\in \mathcal{X}} \frac{\left( \widehat{S}_{\Theta}(\mathbf{u}_{t,x},T_{t,x}) - S_{t,x} \right)^2}{ 2\widehat{V}_{\Theta}(\mathbf{u}_{t,x},T_{t,x}) } + \log ( \widehat{V}_{\Theta}(\mathbf{u}_{t,x},T_{t,x}) ). \label{eq:hgl_loss}
\end{align}
$\widehat{V}_{\Theta}(\mathbf{u}_{t,x},T_{t,x})$ is an estimate of the conditional variance of the subgrid forcing given the low-resolution data, which is also generated by a convolutional neural network. Both loss functions were minimized via batch-based stochastic gradient descent with respect to the network parameters. Additional details about the training procedure are reported in Supplementary Section~\ref{sec:ap:deep-learning}.
 

\subsection{Linear-Inversion Baseline Parameterization}\label{sec:lsrp}
In this section we propose a procedure to estimate subgrid forcing based on partial inversion of the filtering and coarse-graining operations described in Section~\ref{sec:data-generation}. Many subgrid parameterizations are based on an approximate inversion of the filtering operator, achieved for instance via popular inversion methods based on truncated Taylor series expansion and iterative deconvolution \mycite{carati2001modelling, chow2005explicit}. The accuracy of these parameterizations depends on the number of iterations in the inversion procedure. Here we take a different approach, performing a complete (not iterative) inversion. 


Let $L$ represent a linear operator mapping a high-resolution variable defined on the fine grid to its filtered and coarse-grained counterpart. For example, the longitudinal momentum on the coarse grid equals
\begin{align}
u = \cgvar{\filter{\hres{u}}} = L \hres{u},
\end{align}
where $\hres{u}$ is the high-resolution longitudinal momentum (following the notation of Section~\ref{sec:data-generation}). The pseudoinverse $L^{\dagger}$ of the linear operator projects coarse-grid variables onto the fine grid,
\begin{align}
\inv{c} &= L^{\dagger} c, \quad c = \{u,v,T\},\\
\inv{\bm{u}} &= (\inv{u},\inv{v}).
\end{align}
We define the linear-inversion parameterization as the estimate of the subgrid forcing obtained by plugging the fine-grid projection of the corresponding variable into equation \eqref{eq:subgrid-forcing-definition}:  
\begin{align}
\widehat{S}_{\operatorname{inv},c} = \cgvar{\filter{\inv{\bm{u}}}\cdot \nabla \filter{\inv{c}} - \filter{\inv{\bm{u}} \cdot \nabla \inv{c}}}, &\quad c = \{u,v,T\}.
\end{align}
Note that this parameterization completely ignores the high-frequency information suppressed by the coarse-graining operation. In addition, the filtering operation can only be inverted at frequencies where its Fourier transfer function (the set of eigenvalues that defines how the filter attenuates the Fourier harmonics \mycite{grooms2021diffusion}) is nonzero \mycite{stolz1999approximate}. 

The pseudoinverse operator is computed according to the following formula: $L^{\dagger}=L^{T}(LL^{T})^{-1}$, where the mapping $L$ is represented by a sparse matrix and $(LL^{T})^{-1}$ is approximated by a sparse matrix up to numerical precision, as explained in more detail in Supplementary Section~\ref{sec:ap:LB}. 


\section{Experiments and Results}
\label{sec:experiments}

\subsection{Evaluation}
The parameterizations described in Section~\ref{sec:methodology} were evaluated on the held-out test set, containing 10\% of the CM2.6 data. Our main evaluation metric is the coefficient of determination $R^2$. Let $S_{t,x}$ denote the ground-truth subgrid forcing at time $t$ and location $x$, and $\widehat{S}(\mathbf{u}_{t,x},T_{t,x})$ the estimate computed from the low-resolution data $\mathbf{u}_{t,x}$ and $T_{t,x}$. The $R^2$ coefficient is defined as 
\begin{align}
R^2 = 1- \frac{ \sum_{t \in \mathcal{T}_{\operatorname{test}}} \sum_{x 
\in \mathcal{X}} \left( \widehat{S}(\mathbf{u}_{t,x},T_{t,x}) - S_{t,x} \right)^2}{ \sum_{t \in \mathcal{T}_{\operatorname{test}}} \sum_{x 
\in \mathcal{X}} S_{t,x}^2},
\end{align}
where $\mathcal{T}_{\operatorname{test}}$ contains the time indices corresponding to the test set and $\mathcal{X}$ determines the spatial extent over which we evaluate the model.

\subsection{Geographic Extent of the Training Data}
In order to study the effect of the geographic extent of the training data on CNN parameterizations, we train the same CNN models on two training sets with different geographic coverage:
\begin{itemize}
\item The \textbf{4-regions} models were trained on the four regions utilized in \citet{guillaumin2021stochastic}, which are depicted in Supplementary Figure \ref{fig:four-regions}.
\item The \textbf{global} models were trained on the full planet.
\end{itemize}
We trained versions of these models at different resolutions: $0.4^{\circ}$, $0.8^{\circ}$, $1.2^{\circ}$, and $1.6^{\circ}$, obtained by setting the data-coarsening factor $\kappa$ equal to 4, 8, 16 and 32 respectively (see Section~\ref{sec:data-generation}). 

\begin{figure}[t]
\begin{tabularx}{0.95\textwidth} { 
>{\centering\arraybackslash}m{0.001\linewidth} 
>{\centering\arraybackslash}m{0.44\linewidth}  
>{\centering\arraybackslash}m{0.44\linewidth}
>{\centering\arraybackslash}m{0.046\linewidth}}
  &Longitudinal momentum
  & Temperature \\
   \begin{turn}{90}Linear inversion\end{turn} & 
 \includegraphics[width=1\linewidth]{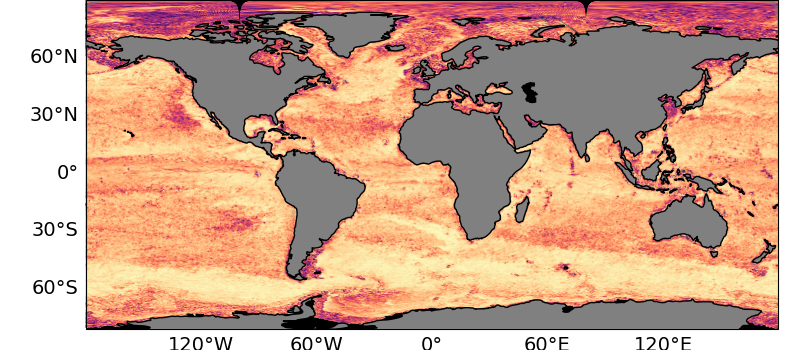}& 
 \includegraphics[width=1\linewidth]{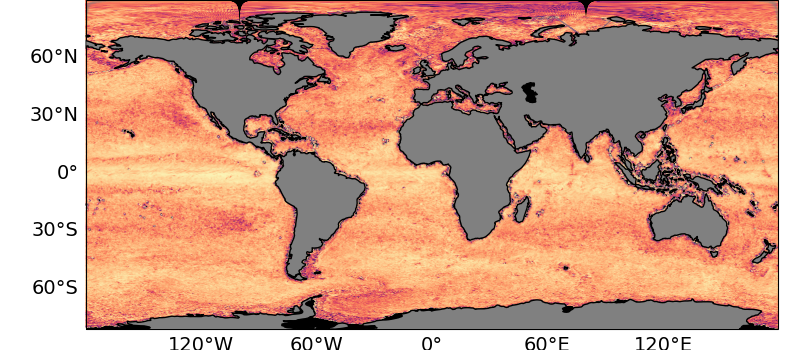}&
 \includegraphics[width=1\linewidth]{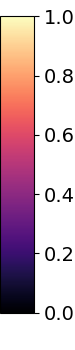}\\
 \begin{turn}{90}\cnnfoureg{}\end{turn} & 
 \includegraphics[width=1\linewidth]{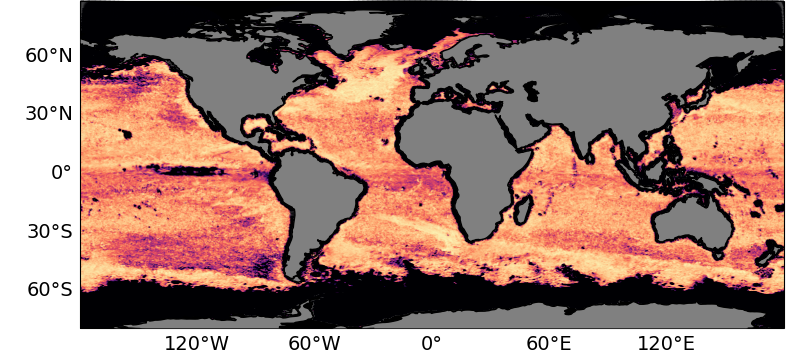}& 
 \includegraphics[width=1\linewidth]{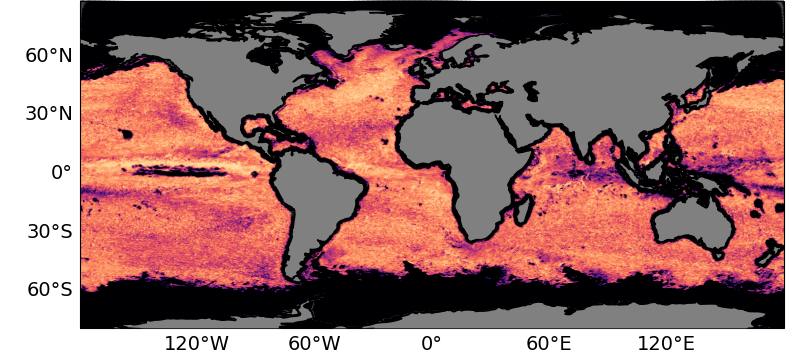}&
 \includegraphics[width=1\linewidth]{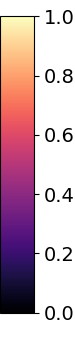}\\
\begin{turn}{90}\cnnglobal{}\end{turn} & 
 \includegraphics[width=1\linewidth]{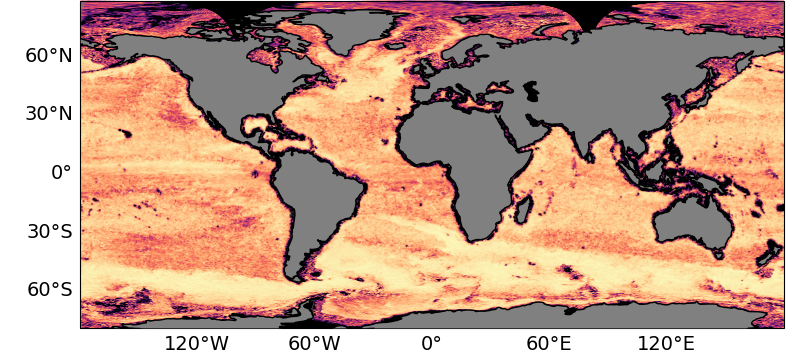}& 
 \includegraphics[width=1\linewidth]{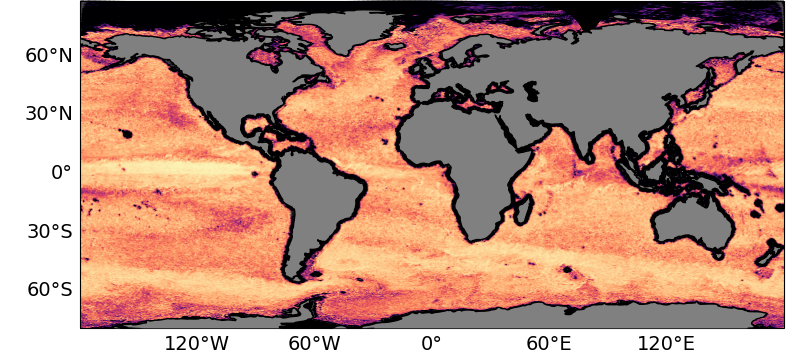}&
 \includegraphics[width=1\linewidth]{images/r2maps/colorbar.png}
\end{tabularx}
\caption{Maps of the $R^2$ coefficient for longitudinal momentum (left column) and temperature (right column) parameterizations computed on a held-out test set. The rows correspond to the linear-inversion parameterization described in Section~\ref{sec:lsrp} (top) and the CNN parameterization described in Section~\ref{sec:data-driven} trained on 4 regions (center) and on the whole planet (bottom). The CNN models were trained by minimizing the MSE loss. The data were processed using General-Circulation-Model filtering and coarse-grained with a factor of 4. The global CNN model trained on the whole planet consistently outperforms the 4-regions model. It also outperforms the linear-inversion baseline, except in at latitudes close to the North Pole. The 4-regions model exhibits poor performance at extreme northern and southern latitudes}
    \label{fig:global-r2-maps}
\end{figure}

\begin{figure}[t]
\begin{tabularx}{0.95\textwidth} { 
>{\centering\arraybackslash}m{0.005\linewidth} 
>{\centering\arraybackslash}m{0.3\linewidth}  
>{\centering\arraybackslash}m{0.3\linewidth}  
>{\centering\arraybackslash}m{0.3\linewidth}  }
&
Longitudinal momentum &
Latitudinal momentum & 
Temperature \\
 \begin{turn}{90} \end{turn} &
 \includegraphics[width=1\linewidth]{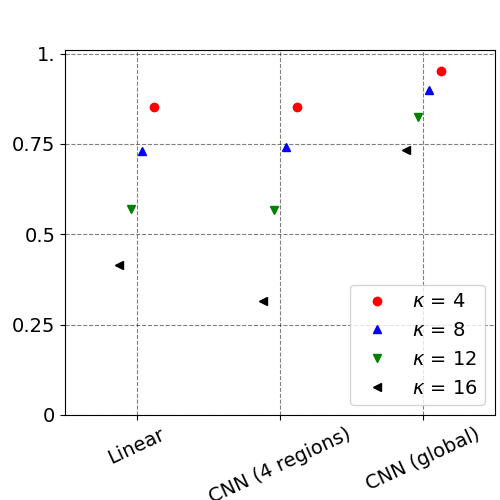}&
 \includegraphics[width=1\linewidth]{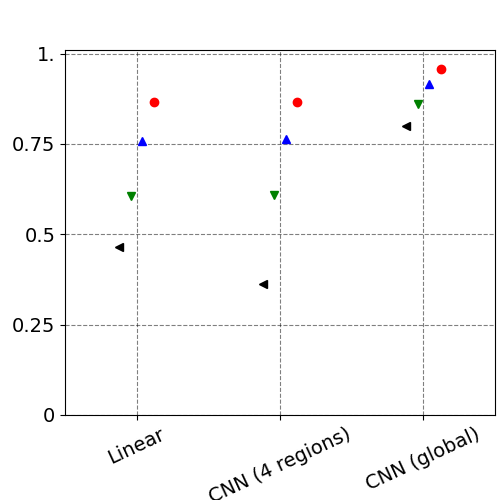}&
 \includegraphics[width=1\linewidth]{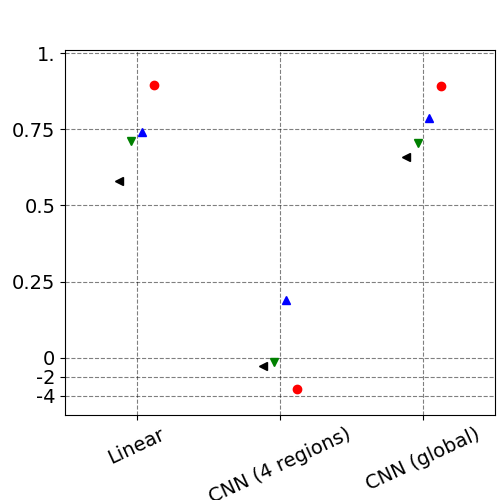}\\
 \begin{turn}{90} +1\% \cotwo{}\end{turn} &
 \includegraphics[width=1\linewidth]{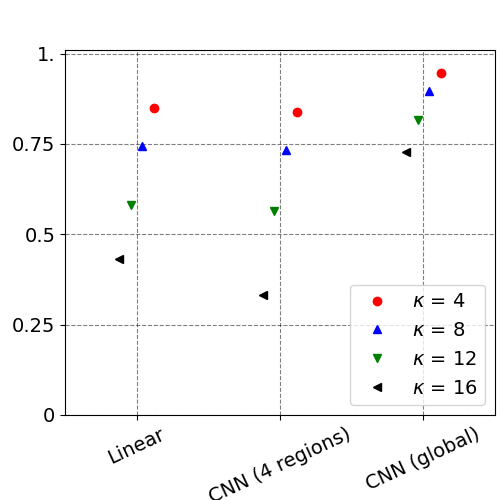}&
 \includegraphics[width=1\linewidth]{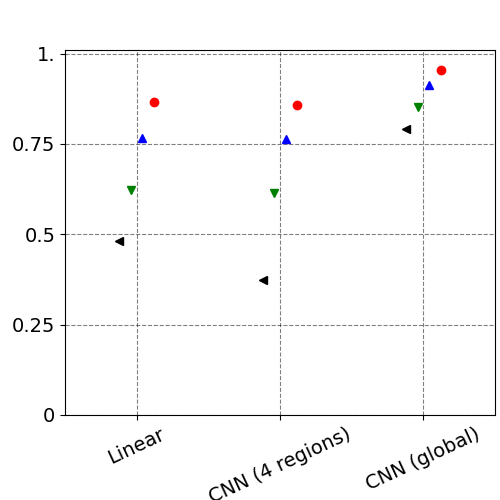}&
 \includegraphics[width=1\linewidth]{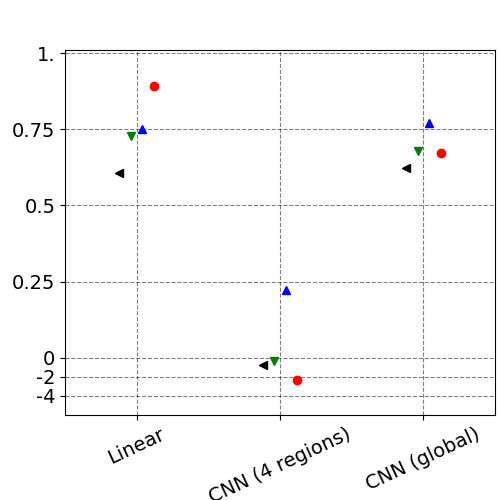}
\end{tabularx}
\caption{
The graphs in the first row show the test $R^2$ coefficient for longitudinal momentum (left), latitudinal momentum (center) and temperature (right) of the linear-inversion parameterization described in Section~\ref{sec:lsrp} and the CNN parameterization described in Section~\ref{sec:data-driven} trained on 4 regions and on the whole planet (global). Results for four different resolutions, corresponding to coarse-graining factors of 4, 8, 12 and 16 are shown. The global model outperforms the 4-regions model and the linear-inversion baseline at all resolutions for both momentum components and temperature.  The graphs in the second row show the same results for data generated at an increased $\text{CO}_2$ level. 
The CNN models were trained by minimizing the MSE loss and the data were processed using General-Circulation-Model filtering. Results for the heteroscedastic Gaussian loss and for Gaussian filtering are shown in Supplementary Figures~\ref{fig:r2-values-gcm} and \ref{fig:r2-values-gauss}, respectively}
    \label{fig:r_squared_MSE}
\end{figure}

Figure~\ref{fig:global-r2-maps} provides a visualization of the performance of both CNN models over the whole globe. The global CNN model trained on the whole planet consistently outperforms the 4-regions model for both momentum components and temperature. The difference in performance is especially pronounced at extreme northern and southern latitudes. The first row of Figure~\ref{fig:r_squared_MSE} reports the aggregated performance of both models for four different resolutions, corresponding to coarse-graining factors of 4, 8, 12 and 16. The global model exhibits strong performance, surpassing the 4-regions model at all resolutions for both momentum and temperature. The difference in performance is particularly marked for temperature, where the 4-regions model performs poorly.   
\begin{figure}[t]
\begin{tabularx}{0.95\textwidth} { 
>{\centering\arraybackslash}m{0.001\linewidth} 
>{\centering\arraybackslash}m{0.22\linewidth}  
>{\centering\arraybackslash}m{0.22\linewidth}
>{\centering\arraybackslash}m{0.22\linewidth}
>{\centering\arraybackslash}m{0.22\linewidth}}
  &$\cgfac = 4$& $\cgfac = 8$ &$\cgfac = 12$ & $\cgfac = 16$\\
 \begin{turn}{90}$\text{S}_u$\end{turn} & 
\includegraphics[width=1\linewidth]
{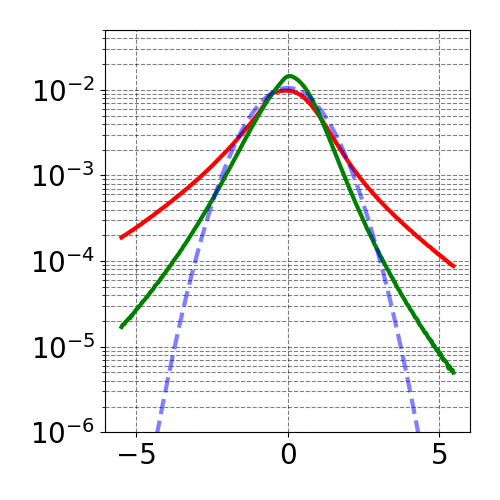}& 
 \includegraphics[width=1\linewidth]
 {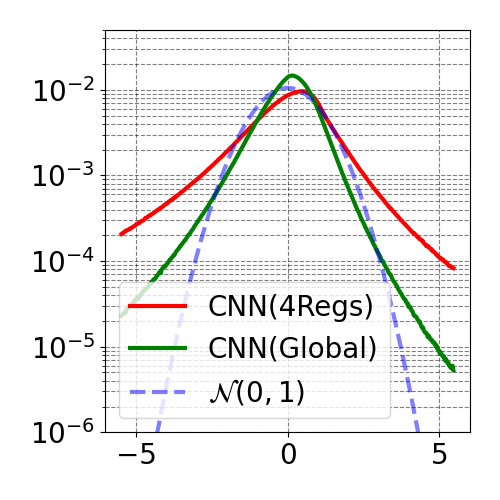}& 
 \includegraphics[width=1\linewidth]
 {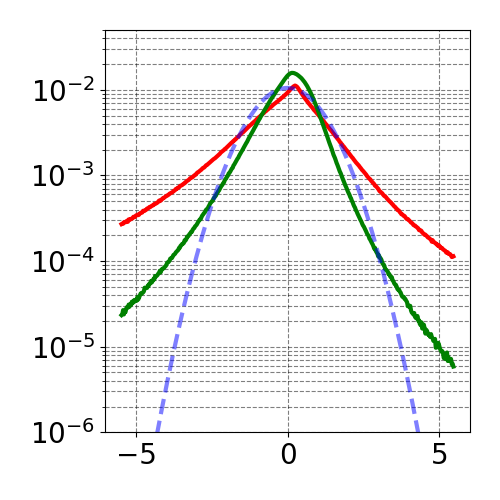}&
 \includegraphics[width=1\linewidth]{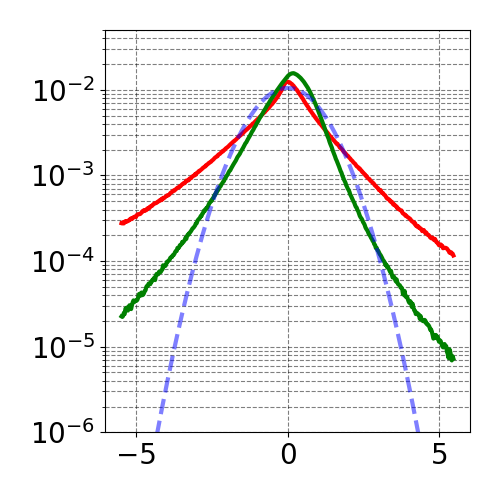}\\
 \begin{turn}{90}$\text{S}_T$\end{turn} & 
\includegraphics[width=1\linewidth]{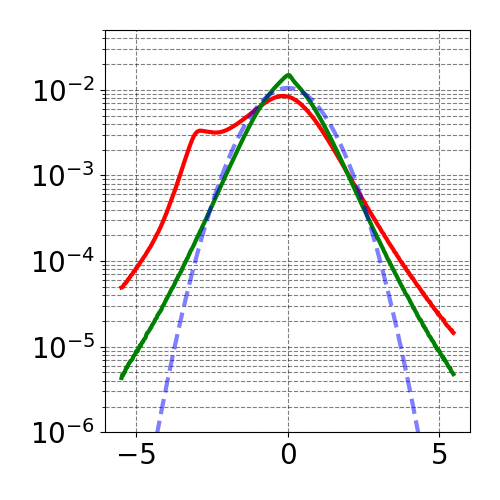}& 
 \includegraphics[width=1\linewidth]{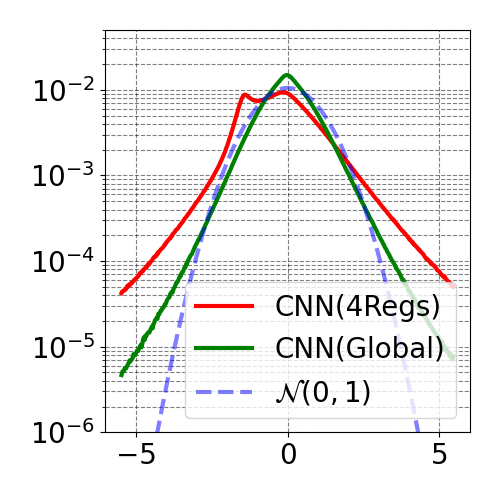}& 
 \includegraphics[width=1\linewidth]{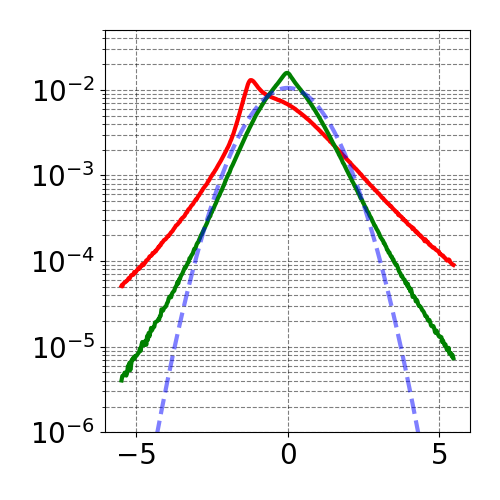}&
 \includegraphics[width=1\linewidth]{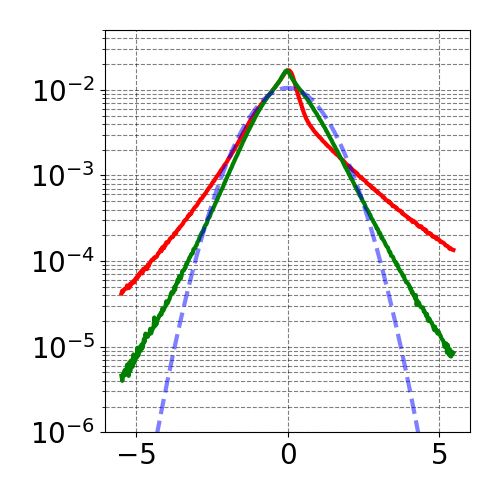}\\
\end{tabularx}
\caption{The graphs show the empirical distributions of the subgrid forcing corresponding to longitudinal momentum (top row) and temperature (bottom row) on the held-out test set, standardized using the conditional mean and conditional variance produced by CNNs trained using the heteroscedastic Gaussian loss~\eqref{eq:hgl_loss}. Results for four different resolutions, corresponding to coarse-graining factors of 4, 8, 12 and 16 are shown. In all cases, the global CNN (green), trained on the entire globe, results in an empirical distribution that more closely resembles a Gaussian with unit variance (blue, dashed) than the 4-regions CNN (red), which indicates that the conditional-variance estimate is improved by increasing the geographic extent of the training data}
    \label{fig:distributions}
\end{figure}

Figure~\ref{fig:distributions} shows that the uncertainty-quantification capabilities of the CNN models are also improved by increasing the geographic extent of the training data. The figure shows the empirical distributions of the subgrid forcing corresponding to longitudinal momentum (top row) and temperature (bottom row) on the held-out test set, standardized using the conditional mean and conditional variance produced by CNNs trained using the heteroscedastic Gaussian loss~\eqref{eq:hgl_loss}. If these conditional estimates are accurate, the standardized distributions should have zero mean and unit variance. The global CNN (green) achieves this to a larger extent than the 4-regions CNN (red). 

\subsection{Comparison with Linear-Inversion Parameterization}

The linear-inversion parameterization presented in Section~\ref{sec:lsrp} estimates the subgrid forcing via a partial linear \emph{inversion} of the coarse-graining operator. It is important to note that this partial linear inversion is nontrivial, because the data-coarsening process is not translation invariant due to coastal boundaries. 

Figure~\ref{fig:global-r2-maps} shows the test performance of this baseline over the whole globe, and compares it to the CNN parameterizations. Close to the North Pole, its performance is superior, and it also outperforms the 4-regions CNN at extreme southern latitudes and some other regions, such as the South Pacific. The first row of Figure~\ref{fig:r_squared_MSE} compares the overall performance of the models at several resolutions. For momentum, the linear baseline is on par with the 4-regions CNN, but inferior to the global CNN. For temperature, it is on par with the global CNN, and superior to the 4-regions CNN. We conclude that partial linear inversion provides a strong baseline, and that the global CNN is able to learn high-resolution structure that is inaccessible via linear inversion.

\subsection{Generalization Across \cotwo{} Levels and Depth}
The second row of Figure~\ref{fig:r_squared_MSE} reports the overall performance of CNN models trained at pre-industrial \cotwo{} levels, when tested on a forced simulation with a $1\%$ increase in \cotwo{} level per year, until the level doubles after 70 years \mycite{griffies2015handbook}. Overall, we observe that the models preserve similar performance levels, indicating that they are able to generalize robustly across different \cotwo{} levels. Generalization for momentum is better than for temperature. This may be partially explained by the fact that the \cotwo{} increase produces a greater distribution shift for temperature than for momentum, as shown in Supplementary Figure~\ref{fig:forcing-distributions}.


\begin{figure}[tp!]
\begin{tabularx}{\textwidth} { 
>{\centering\arraybackslash}m{0.47\linewidth}  
>{\centering\arraybackslash}m{0.47\linewidth}}
 Longitudinal momentum
 & Temperature 
 \end{tabularx}\\\includegraphics[width=1\linewidth]{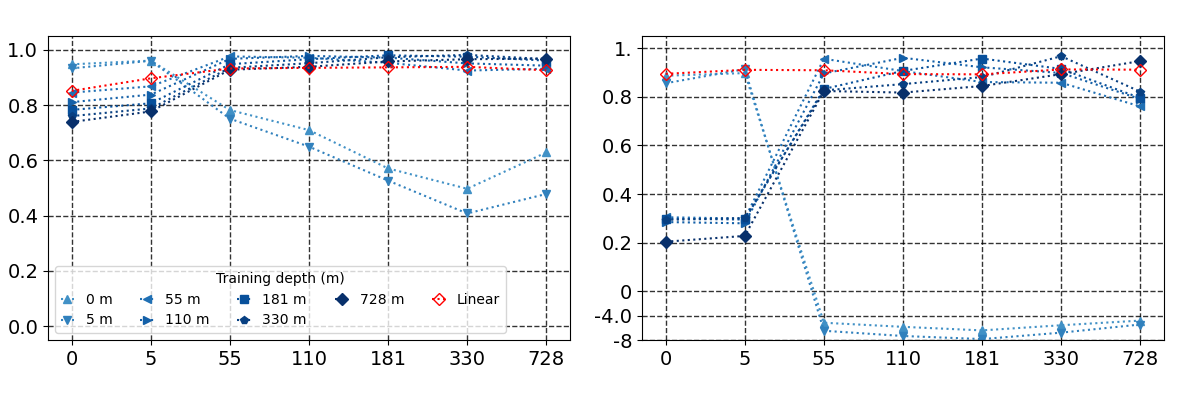}\\ \includegraphics[width=1\linewidth]{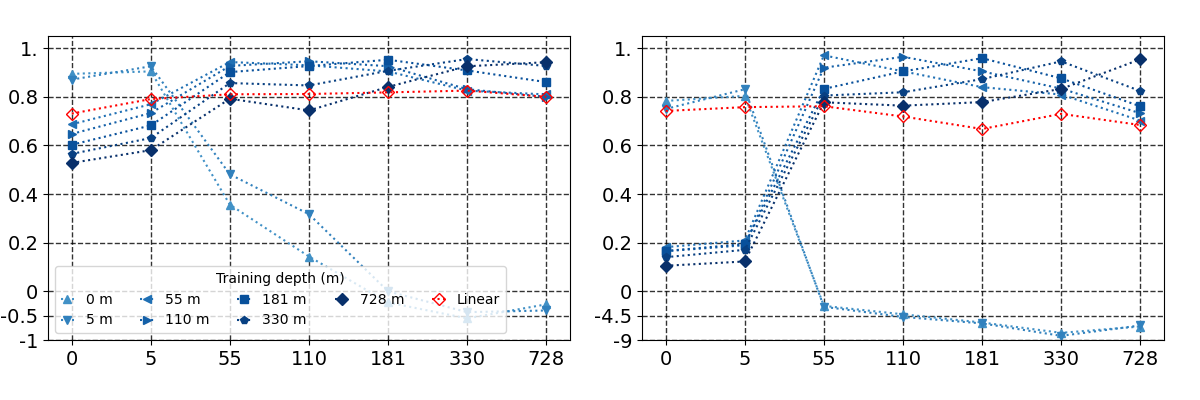}\\
\includegraphics[width=1\linewidth]{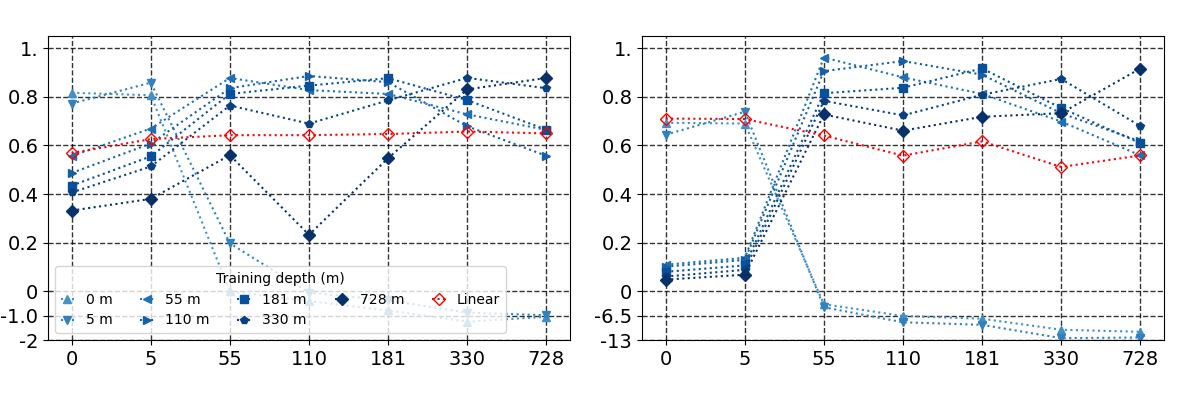}
\caption{The graphs show the test $R^2$ coefficient at different depths for the longitudinal momentum (left) and temperature (right) of the linear-inversion parameterization (in red) described in Section~\ref{sec:lsrp} and the CNN parameterization (in blue) described in Section~\ref{sec:data-driven} trained on the whole planet (global). The different markers indicate the depth at which the corresponding CNN model was trained. Models trained at shallow depth (surface and 5 m) do not generalize well to greater depth. Conversely, models trained at greater depth do not generalize well to the surface. The CNN models were trained by minimizing the MSE loss. The data were processed using General-Circulation-Model filtering and coarse-grained with a factors of 4 (top), 8 (center) and 12 (bottom)} 

    \label{fig:sigma-8-r-square-depths}
\end{figure}

In addition, we investigated the performance of CNN-parameterizations at different depths, utilizing the subsurface training and test sets described in Section~\ref{sec:data-driven}. We trained separate CNN models on the whole planet at six different depths: 5 m, 55 m, 110 m, 181 m, 330 m, and 728 m. We tested these models and the surface model at all depths (including the surface). Figure~\ref{fig:sigma-8-r-square-depths} shows the results. The surface model generalizes well to 5 m and vice versa, but both models do not generalize robustly to greater depths. Conversely, models trained at depths 55 m and beyond do not generalize robustly to the surface or to 5 m, but do generalize to the remaining depths. This indicates that the models learn very different structure at the surface and near-surface, with respect to greater depths. The linear-inversion parameterization, which is agnostic to physical structures, shows uniform performance at all depths. However, at most depths the linear-inversion model is less accurate than the CNN-parameterization trained at the same depths.

\subsection{Input Stencil}
\label{sec:input_stencil}
\begin{figure}[t]
\begin{tabularx}{\textwidth} { 
>{\centering\arraybackslash}m{0.47\linewidth}  
>{\centering\arraybackslash}m{0.47\linewidth}}
 Longitudinal momentum
 & Temperature 
 \\
\includegraphics[width = 1\linewidth]{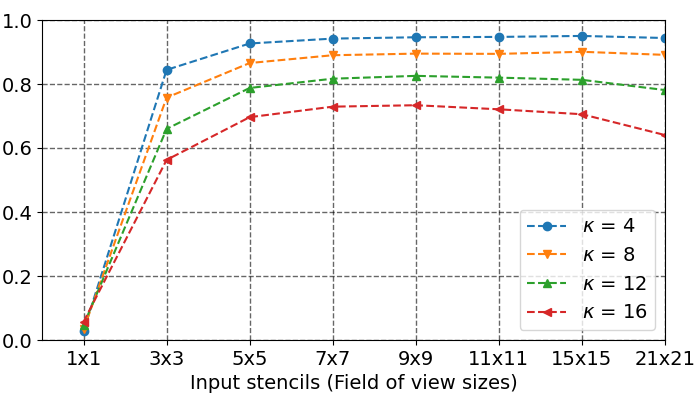} &
\includegraphics[width = 1\linewidth]{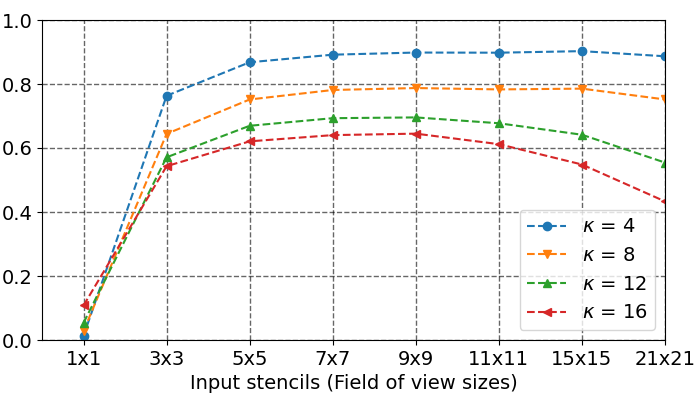}\\
\end{tabularx}
\caption{
The graphs plot the test $R^2$ coefficient for the longitudinal momentum (left) and temperature (right) of CNN parameterizations with different input stencils or fields of view (in pixels). Results for four different resolutions, corresponding to coarse-graining factors of 4, 8, 12 and 16 are shown. The CNN models were trained by minimizing the MSE loss. The data were processed using General-Circulation-Model filtering. At all resolutions performance saturates when the stencil size reaches $7\times7$.
}
    \label{fig:stencil-sweep}
\end{figure}

A key consideration for the practical deployment of CNN-based parameterizations is their computational cost, in particular in the absence of GPUs \citep{zhang2023implementation}. This cost depends on the input stencil or field of view of the CNN, which is the spatial extent utilized by the model to produce an estimate at each location. To determine the influence of the input stencil on the performance of these models, we trained different models with the same number of parameters, but different input stencils. This was achieved by reducing the size of the convolutional filters in each layer, while simultaneously increasing the number of filters to preserve the overall number of parameters.  

Figure~\ref{fig:stencil-sweep} shows the test $R^2$ coefficient for the longitudinal momentum (left) and temperature (right) of CNN parameterizations with different input stencils. At all resolutions performance saturates when the stencil size reaches $7\times7$. This is a considerable reduction from the $21\times 21$ input stencils of previous CNNs~\mycite{guillaumin2021stochastic}. 

\begin{figure}[t]
\begin{tabularx}{\textwidth} { 
>{\centering\arraybackslash}m{0.3\linewidth}  
>{\centering\arraybackslash}m{0.3\linewidth}
>{\centering\arraybackslash}m{0.3\linewidth}}
  Longitudinal momentum & Latitudinal momentum & Temperature \\
\includegraphics[width=1\linewidth]{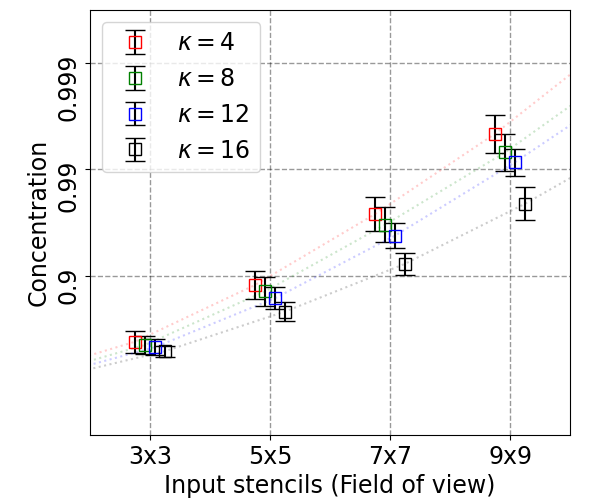} &
  \includegraphics[width=1\linewidth]{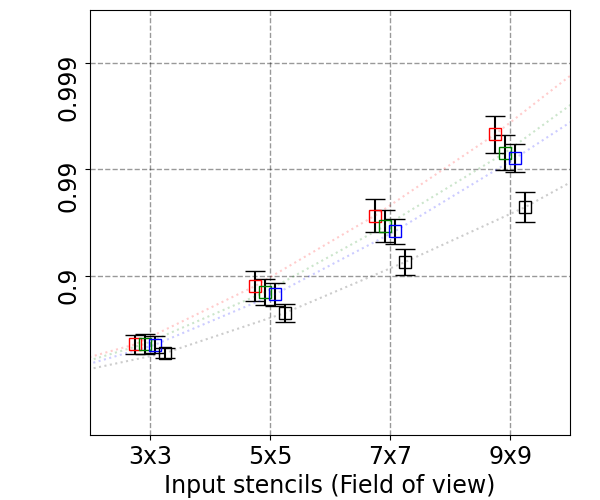} &
  \includegraphics[width=1\linewidth]{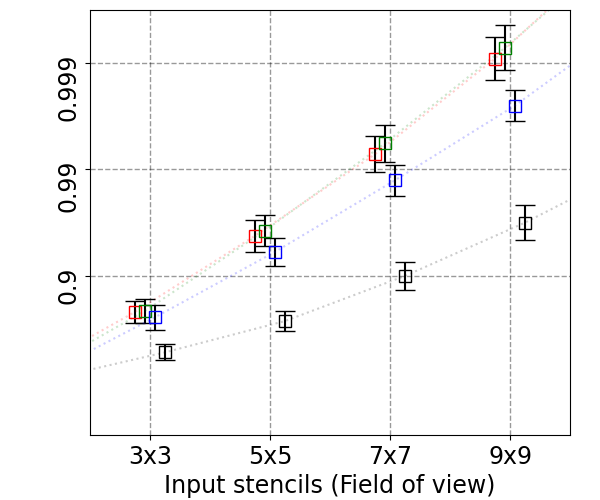}  
\end{tabularx}
\caption{
The graphs plot the median concentration of the gradient energy of CNN parameterizations for longitudinal momentum (left), latitudinal momentum (center) and temperature (right), along with error bars indicating the 10$\%$ and 90$\%$ quantiles. The concentration is measured as the fraction of the sum of squared gradient amplitude within an input stencil or field of view of a certain size.  Results for four different resolutions, corresponding to coarse-graining factors of 4, 8, 12 and 16 are shown. The CNN models were trained by minimizing the MSE loss. The data were processed using General-Circulation-Model filtering. At all resolutions 90$\%$ or more gradient energy is concentrated within a stencil size of $7\times7$}
    \label{fig:saliency}
\end{figure}

In order to test the hypothesis that CNN parameterizations primarily leverage information that closely surrounds the output location, we performed a gradient-based analysis. The gradient of neural-network outputs with respect to their input is widely used to interpret the functions learned by these models~\mycite{simonyan2013deep,mohan2019robust,ross2023benchmarking}. It quantifies the sensitivity of the network output to small changes in each input pixel. We measure the concentration of the gradient by computing the fraction of the sum of squared gradient amplitude that lies within a certain input stencil. Figure~\ref{fig:saliency} shows the median concentration of the gradient energy of CNN parameterizations for longitudinal momentum (left), latitudinal momentum (center) and temperature (right). At all resolutions, 90$\%$ or more gradient energy is concentrated within a stencil size of $7\times7$, which supports our hypothesis that the network primarily leverages a small input region to generate its output.

\section{Discussion and Outlook}\label{sec:discussion-outlook}

This study revealed several insights regarding data-driven parameterizations for mesoscale eddies based on convolutional neural networks, which may be useful for the design and implementation of these models in other situations. First, we observed that the geographic extent represented in the training data has a significant influence on model performance. Second, we compared CNN parameterizations with a linear baseline, based on partial inversion of the coarse-graining operation typically used to generate training data for data-driven parameterizations. The CNN parameterizations mostly outperform the baseline, which suggests that the CNNs are able to learn nonlinear physical structure. 
Third, we evaluated the generalization ability of CNN parameterizations at different \cotwo{} levels and ocean depths, finding that they are able to generalize to increased \cotwo{} levels, but not from surface to subsurface depths beyond 50 m (or vice versa). It is possible that  surface atmospheric forcing has a strong influence in the upper ocean, which is not present below the surface, creating a distribution shift that leads to a lack of generalization. 
Fourth, we determined that CNN parameterizations mainly exploit a relatively small surrounding region of their input to produce their outputs. These insights were consistent for both momentum and temperature parameterizations across different resolutions.  

Our study focused on offline metrics, which quantify to what extent parameterizations are able to approximate the missing subgrid forcing in coarse-grid climate simulations. A crucial direction for future research is to evaluate parameterizations in an online setting, once incorporated in realistic climate models (see~\mycite{zhang2023implementation, perezhogin2023implementation} for recent progress in this direction).  

\begin{Backmatter}


\paragraph{Funding Statement}
This project is supported by Schmidt Sciences, LLC. C.F.G was partially supported by NSF DMS Grant 2009752.

\paragraph{Competing Interests}
The authors declare none.

\paragraph{Data Availability Statement}
The authors downloaded the simulation's ocean surface velocities from a Pangeo data catalog at \url{https://raw.githubusercontent.com/pangeo-data/pangeo-datastore/master/intake-catalogs/ocean/GFDL_CM2.6.yaml} made publicly available by the Geophysical Fluid Laboratory.

\paragraph{Ethical Standards}
The research meets all ethical guidelines, including adherence to the legal requirements of the study country.

\paragraph{Author Contributions}
Analysis and investigation: C.G.; Conceptualization: A.A., C.F.G., and L.Z.; Data curation: C.G., P.P., C.Z., and A.S.; Data visualization: C.G., Methodology: C.G., M.L., C.F.G., and L.Z.; Supervision: A.A., C.F.G., and L.Z.; Writing original draft: C.G., C.F.G., P.P., and L.Z.; Writing—reviewing and editing: C.G., P.P., C.Z., A.A., C.F.G., and L.Z. All authors approved the final submitted draft.

\printbibliography







\end{Backmatter}

\begin{appendix}
\clearpage

\section{Neural-Network Implementation Details}\label{sec:ap:deep-learning}
Following \mycite{guillaumin2021stochastic} the convolutional neural networks used in this study consist of eight convolutional layers interleaved with ReLU nonlinearities. Table~\ref{tab:base-arch} provides a detailed description of the convolutional layers used for the different coarse-graining factors. We incorporate a batch-normalization layer~\mycite{ioffe2015batch} after each intermediate layer. For the models trained with the heteroscedastic Gaussian, the output layer has multiple channels, associated with the conditional mean and conditional variance estimate. The latter estimate is generated in the form of the inverse conditional standard deviation, which is constrained to be nonnegative by a softplus activation function, as in \mycite{guillaumin2021stochastic}. 

\begin{table}[t]
    \centering
    \begin{tabular}{ |p{1.3cm}||p{1.4cm}|p{1.4cm}|p{1cm}|p{1cm}|p{1cm}|p{1cm}|p{1cm}|p{1.1cm}|}
\hline
  & layer-1 & layer-2 & layer-3 & layer-4 & layer-5 & layer-6 & layer-7 & layer-8  \\
 \hline
 $\cgfac = $ 4 &  &  & & &  & & &  \\
 \hline
 width & (2,3)x128 & 128x64 & 64x 32 &32x 32& 32x32 &32x32 &32x32 &32x(4,6)  \\
 \hline
kernel size & 5x5 & 5x5 & 3x3 & 3x3 & 3x3 & 3x3 & 3x3 & 3x3 \\
 \hline
 \hline
 $\cgfac = $ 8 &  &  & & &  & & &  \\
 \hline
 width & (2,3)x208 & 208x104 & 104x52 & 52x52 & 52x52 & 52x52 & 52x52 &52x(4,6)  \\
  \hline
kernel size & 3x3 & 3x3 & 2x2 & 2x2 & 2x2 & 2x2 & 2x2 & 2x2 \\
 \hline
 \hline
 $\cgfac = $ 12,16 &  &  & & &  & & &  \\
 \hline
 width & (2,3)x279 & 279x140 & 140x70 & 70x70 & 70x70 & 70x70 & 70x70 &70x(4,6) \\
  \hline
kernel size & 2x2 & 2x2 & 2x2 & 2x2 & 2x2 & 2x2 & 1x1 & 1x1 \\
 \hline 
\end{tabular} \caption{Structure of the convolutional layers of the CNN architectures used for each coarse-graining factor $\cgfac$}
    \label{tab:base-arch}
\end{table}

During training, we use a variable learning rate scheduler called \texttt{ReduceLROnPlateau}, available in \texttt{pytorch}. Starting from an initial learning rate (equal to 0.01), the scheduler checks if the validation error has decreased at the end of each epoch. If this is not the case, over a certain number of epochs (equal to 3), the learning rate is decreased. The process continues until the learning rate reaches below a certain threshold (equal to $10^{-7}$). The training mini-batches contain four frames for a coarse-graining factor of 4, as in \mycite{guillaumin2021stochastic}. At larger factors, the minibatch sizes are increased to preserve the number of grid points.

In Section~\ref{sec:input_stencil} we report results for different CNNs with the same number of parameters, but different input stencils. This was achieved by reducing the size of the convolutional filters in each layer, while simultaneously increasing the number of filters to preserve the overall number of parameters. This was achieved by first shrinking the kernel sizes to achieve a certain stencil size, and then increasing the number of channels.

\begin{table}[t]
    \centering
    \begin{tabular}{ |p{1.6cm}||p{.95cm}|p{.95cm}|p{.95cm}|p{.95cm}|p{.95cm}|p{.95cm}|p{.95cm}|p{.95cm}|p{.95cm}|}
\hline
  & lyr-1 & lyr-2 & lyr-3 & lyr-4 & lyr-5 & lyr-6 & lyr-7 & lyr-8 & stencil  \\
 \hline
 width & 128 & 64 & 32 & 32& 32 & 32 & 32 & 32 &   \\
ker. size & 5 & 5 & 3 & 3 & 3 & 3 & 3 & 3 & 21x21\\
 \hline
width & 186 & 93 & 46 & 46& 46 & 46 & 46 & 46 &   \\
ker. size & 3 & 3 & 3 & 3 & 3 & 3 & 2 & 2 & 15x15\\
\hline
width & 208 & 104 & 52 & 52& 52 & 52 & 52 & 52 &   \\
ker. size & 3 & 3 & 2 & 2 & 2 & 2 & 2 & 2 & 11x11\\
\hline
width & 271 & 136 & 68 & 68& 68 & 68 & 68 & 68 &   \\
ker. size & 2 & 2 & 2 & 2 & 2 & 2 & 2 & 2 & 9x9\\
\hline
width & 279 & 140 & 70 & 70& 70 & 70 & 70 & 70 &   \\
ker. size & 2 & 2 & 2 & 2 & 2 & 2 & 1 & 1 & 7x7\\
\hline
width & 296 & 148 & 74 & 74& 74 & 74 & 74 & 74 &   \\
ker. size & 2 & 2 & 2 & 2 & 1 & 1 &1 & 1 & 5x5\\
\hline
width & 327 & 164 & 82 & 82& 82 & 82 & 82 & 82 &   \\
ker. size & 2 & 2 & 1 & 1 & 1 & 1 & 1 & 1 & 3x3\\
\hline
width & 528 & 264 & 132 & 132& 132 & 132 & 132 & 132 &   \\
ker. size & 1 & 1 & 1 & 1 & 1 & 1 & 1 & 1 & 1x1\\
\hline
\end{tabular} 
\caption{Across the study, we use adapt the original 8 layer CNN architecture to various stencil sizes. The stencil size refers to the total input window size of the CNNs. Each row corresponds to a CNN model. Each convolutional layer has a number of output channels (width) and filter size (ker. size). All kernels are square shaped and the total input sizes (stencil) are provided in the last column.}
    \label{tab:stencil-arch}
\end{table}

\section{Linear Baseline Details}\label{sec:ap:LB}

Section~\ref{sec:lsrp} proposes a procedure to estimate subgrid forcing based on partial inversion of the filtering and coarse-graining operations described in Section~\ref{sec:data-generation}. The procedure involves computing the pseudoinverse of the linear operator $L$ that  maps the high-resolution variables defined on the fine grid to their filtered and coarse-grained counterparts. 

We first reshape two-dimensional arrays of velocities or temperature into one-dimensional vectors and exclude the land points. This allows to represent operator $L$ with a sparse matrix. The pseudoinverse is given by the formula $L^{\dagger}=L^{T}(LL^{T})^{-1}$,  
where the inverse matrix ($(LL^{T})^{-1}$) is computed via hierarchical matrix factorization following \citet{borm2003introduction}. One factorization step divides any matrix $M$ into 4 blocks as follows:
\begin{equation}
    M = \begin{pmatrix}
            A & B \\ C& D
        \end{pmatrix},
\end{equation}
where each block is half the size of the original matrix. The block matrix can be formally inverted as follows:
\begin{equation}
   M^{-1} = \begin{pmatrix}
            (M/D)^{-1} & -(M/D)^{-1}BD^{-1}\\
            -D^{-1}C(M/D)^{-1} & D^{-1} + D^{-1}C(M/D)^{-1}BD^{-1}
        \end{pmatrix},
\end{equation}
where $M/D = A - BD^{-1}C$ is the Schur complement. The procedure is applied recursively to the matrix $L L^T$ until the individual matrices can be directly inverted. An essential ingredient of our approach is the approximation of the inverse matrix ($(L L^T)^{-1}$) by a sparse matrix. Sparsity of the inverse matrix is achieved by neglecting matrix elements that are below the numerical precision during each factorization step. 

\section{Additional Figures}

\begin{figure}[t]
    \centering
    \includegraphics[width=10cm]{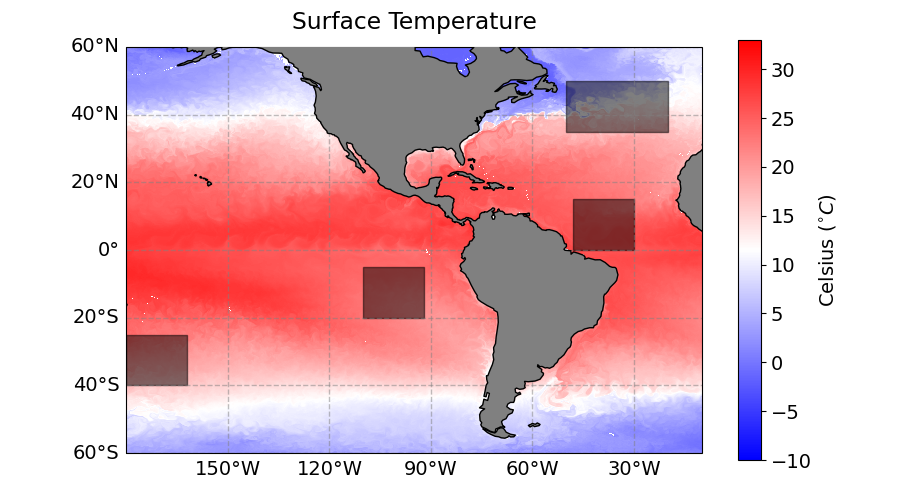}
    \label{fig:four-regions}
\caption{The shaded regions indicate the geographic locations used to train the 4-regions CNN model following \mycite{guillaumin2021stochastic}}
\end{figure}

\begin{figure}[t]
\begin{tabularx}{0.95\textwidth} { 
>{\centering\arraybackslash}m{0.005\linewidth} 
>{\centering\arraybackslash}m{0.4\linewidth}
>{\centering\arraybackslash}m{0.4\linewidth}
>{\centering\arraybackslash}m{0.04\linewidth}}
$\text{R}^2_T$ & Heteroscedastic & MSE  &
\\
 \begin{turn}{90}$\cgfac = 4$\end{turn} &
 \includegraphics[width=1\linewidth]{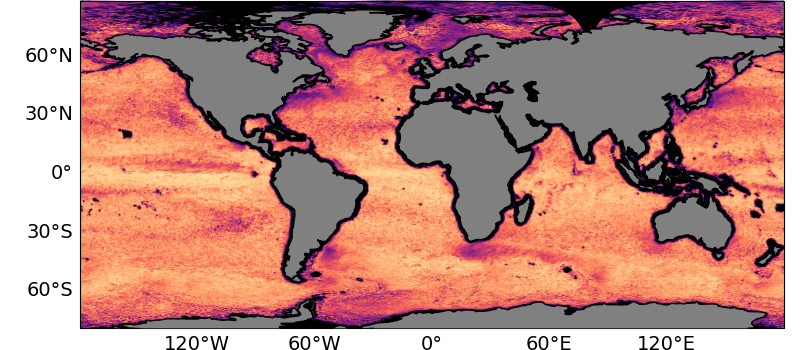}&
  \includegraphics[width=1\linewidth]{images/r2maps/flt_gcm_lss_MSE_sgm_4_flt_gcm_c2_0p00_global_Stemp.png} &
    \includegraphics[width=1\linewidth]{images/r2maps/colorbar.png} 
  \\
\begin{turn}{90}$\cgfac = 8$\end{turn} &
 \includegraphics[width=1\linewidth]{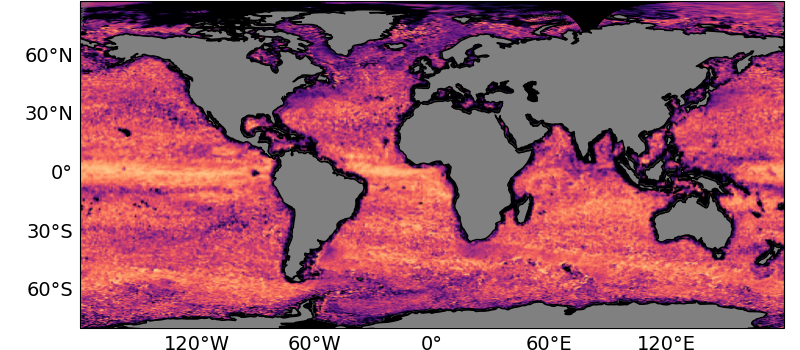}&
  \includegraphics[width=1\linewidth]{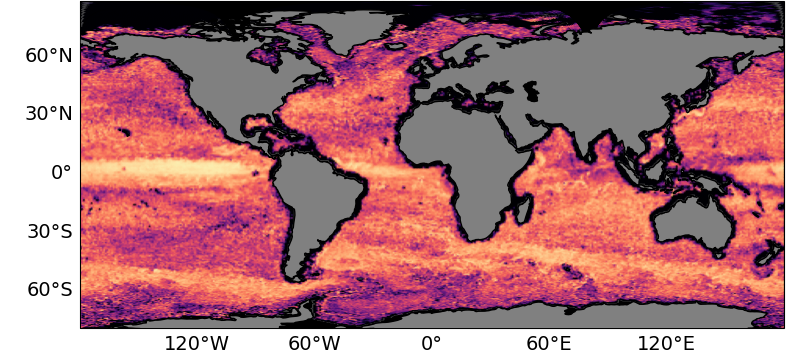} &
    \includegraphics[width=1\linewidth]{images/r2maps/colorbar.png} 
  \\
\begin{turn}{90}$\cgfac = 12$\end{turn} &
 \includegraphics[width=1\linewidth]{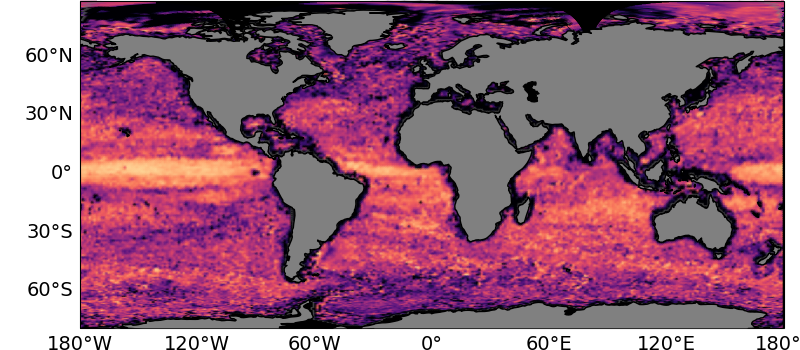}&
  \includegraphics[width=1\linewidth]{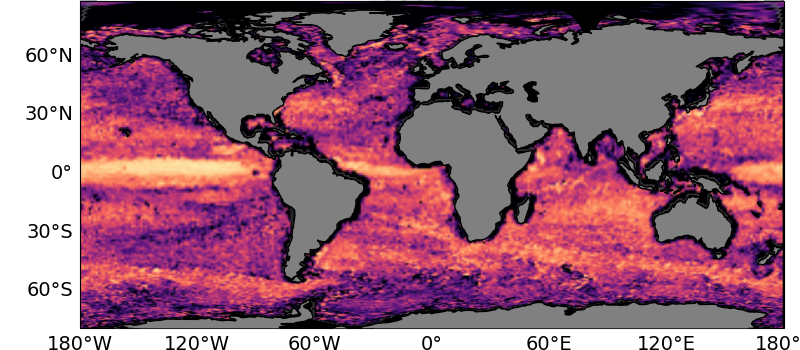} &
    \includegraphics[width=1\linewidth]{images/r2maps/colorbar.png} 
  \\
\begin{turn}{90}$\cgfac = 16$\end{turn} &
 \includegraphics[width=1\linewidth]{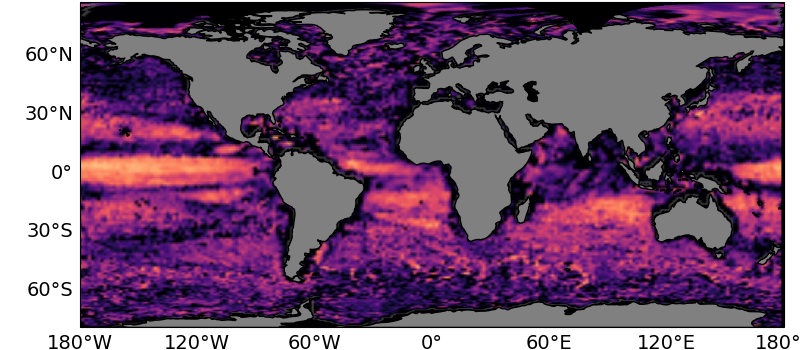}&
  \includegraphics[width=1\linewidth]{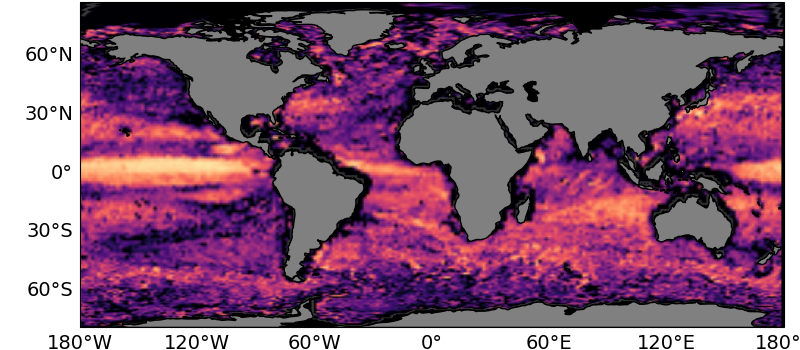} &
    \includegraphics[width=1\linewidth]{images/r2maps/colorbar.png} 
  \\
\end{tabularx}
\caption{Maps of the test $R^2$ coefficient for temperature of the CNN parameterization trained on the whole globe by minimizing the heteroscedastic Gaussian loss (left) and MSE (right). The data were processed using General-Circulation-Model filtering at coarse-grained at different factors, indicated by the value of $\kappa$ in each row}
    \label{fig:coordinatewise-r2-global-all-sigmas}
\end{figure}

\begin{figure}[t]
\begin{tabularx}{0.95\textwidth} { 
>{\centering\arraybackslash}m{0.005\linewidth} 
>{\centering\arraybackslash}m{0.3\linewidth}  
>{\centering\arraybackslash}m{0.3\linewidth}  
>{\centering\arraybackslash}m{0.3\linewidth}  }
&
Longitudinal momentum &
Latitudinal momentum & 
Temperature \\
   \begin{turn}{90} HGL loss \end{turn} &
   \includegraphics[width = \linewidth]{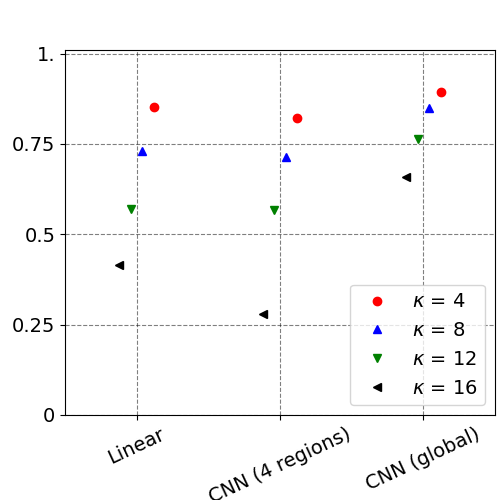}&
  \includegraphics[width = \linewidth]{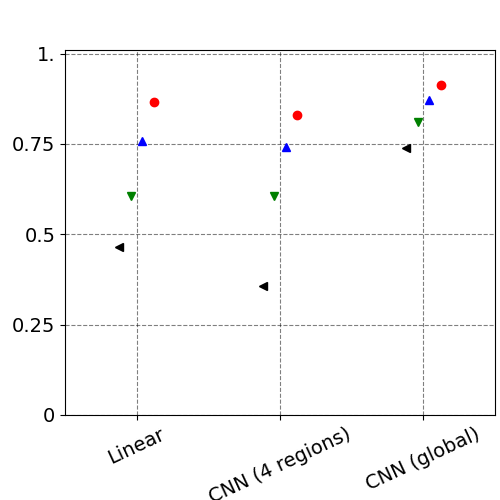}&
   \includegraphics[width = \linewidth]{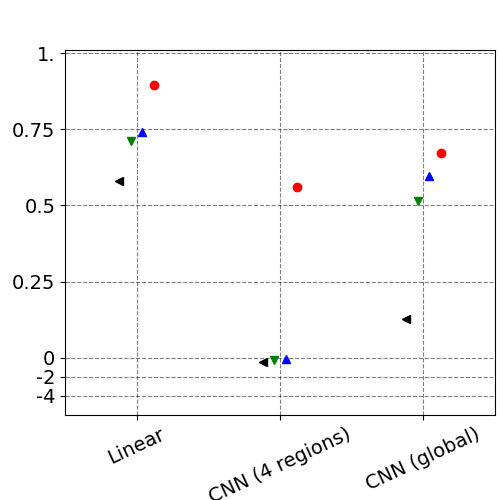} \\
  \begin{turn}{90} HGL loss (+1\% \cotwo{})\end{turn} &
\includegraphics[width=1\linewidth]{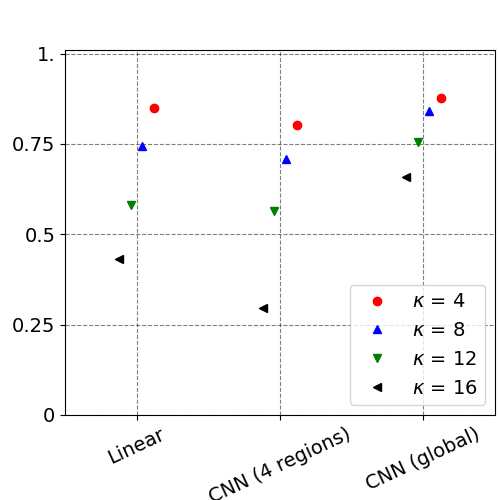}&
 \includegraphics[width=1\linewidth]{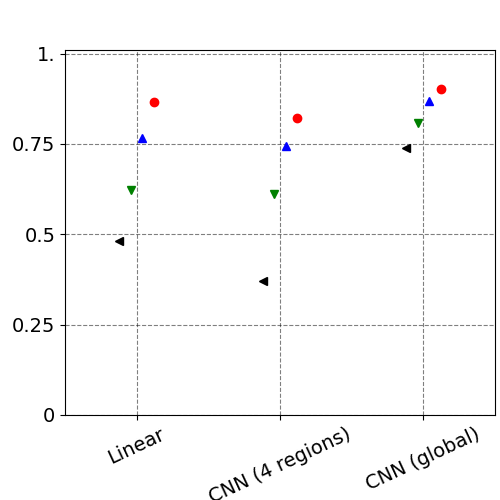}&
 \includegraphics[width=1\linewidth]{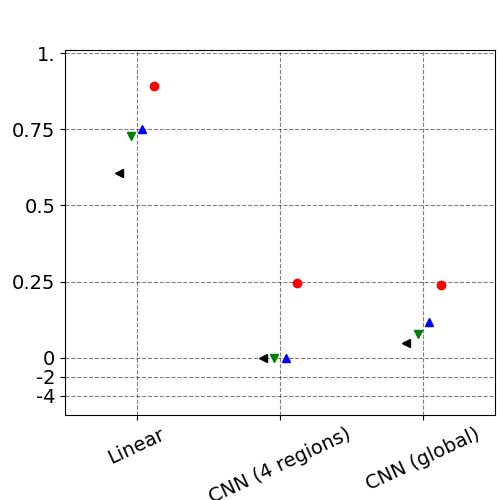}\\
\end{tabularx}
\caption{The graphs in the first row show the test $R^2$ coefficient for longitudinal momentum (left), latitudinal momentum (center) and temperature (right) of the linear-inversion parameterization described in Section~\ref{sec:lsrp} and the CNN parameterization described in Section~\ref{sec:data-driven} trained on 4 regions and on the whole planet (global). Results for four different resolutions, corresponding to coarse-graining factors of 4, 8, 12 and 16 are shown. The graphs in the second row show the same results for data generated at an increased $\text{CO}_2$ level. 
The CNN models were trained by minimizing the heteroscedastic Gaussian loss and the data were processed using General-Circulation-Model filtering}
    \label{fig:r2-values-gcm}
\end{figure}

\begin{figure}[t]
\begin{tabularx}{0.95\textwidth} { 
>{\centering\arraybackslash}m{0.005\linewidth} 
>{\centering\arraybackslash}m{0.3\linewidth}  
>{\centering\arraybackslash}m{0.3\linewidth}  
>{\centering\arraybackslash}m{0.3\linewidth}  }
&
Longitudinal momentum &
Latitudinal momentum & 
Temperature \\
 \begin{turn}{90}MSE loss \end{turn} &
 \includegraphics[width=1\linewidth]{images/basic/r2_co2_0p00_filtering_gaussian_lossfun_MSE_training_filtering_gaussian_Su_r2.png}&
 \includegraphics[width=1\linewidth]{images/basic/r2_co2_0p00_filtering_gaussian_lossfun_MSE_training_filtering_gaussian_Sv_r2.png}&
 \includegraphics[width=1\linewidth]{images/basic/r2_co2_0p00_filtering_gaussian_lossfun_MSE_training_filtering_gaussian_Stemp_r2.png}\\
 \begin{turn}{90}MSE loss (+1\% \cotwo{})\end{turn} &
 \includegraphics[width=1\linewidth]{images/basic/r2_co2_0p01_filtering_gaussian_lossfun_MSE_training_filtering_gaussian_Su_r2.png}&
 \includegraphics[width=1\linewidth]{images/basic/r2_co2_0p01_filtering_gaussian_lossfun_MSE_training_filtering_gaussian_Sv_r2.png}&
 \includegraphics[width=1\linewidth]{images/basic/r2_co2_0p01_filtering_gaussian_lossfun_MSE_training_filtering_gaussian_Stemp_r2.png}\\
 \begin{turn}{90}HGL loss\end{turn} &
\includegraphics[width=1\linewidth]{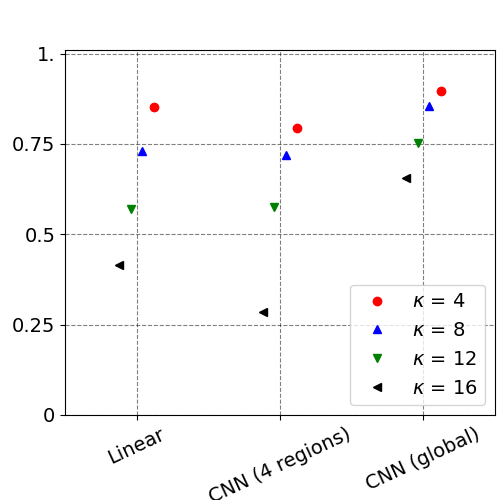}&
\includegraphics[width=1\linewidth]{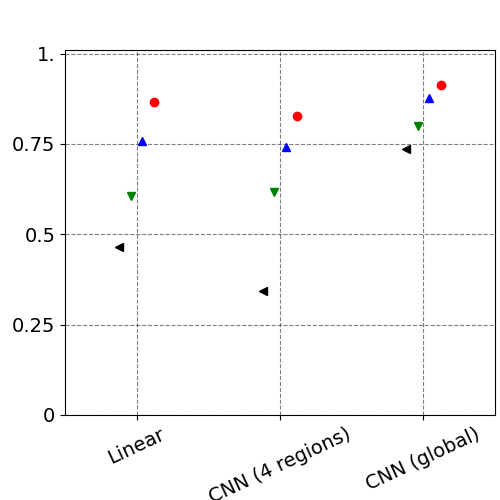}&
\includegraphics[width=1\linewidth]{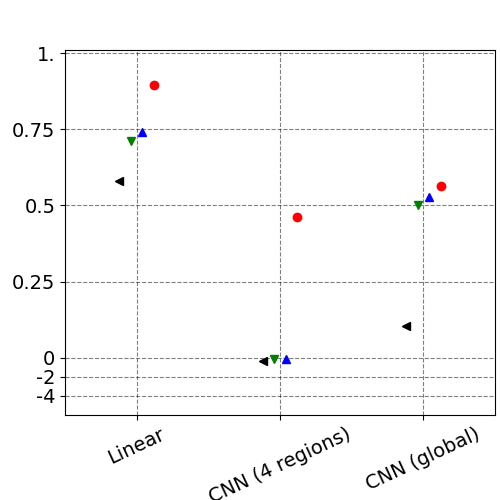}
 \\
\includegraphics[width=1\linewidth]{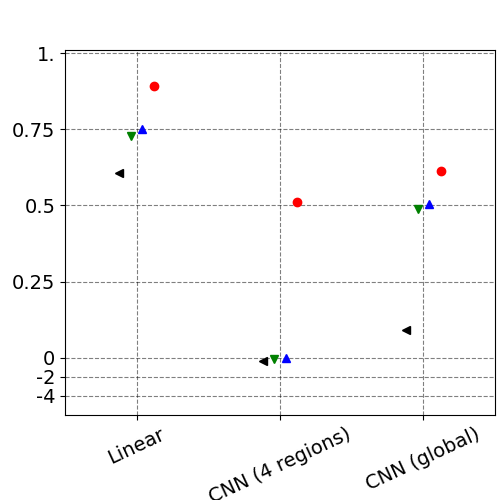}\\
   \begin{turn}{90}HGL loss (+1\% \cotwo{})\end{turn} &
\includegraphics[width=1\linewidth]{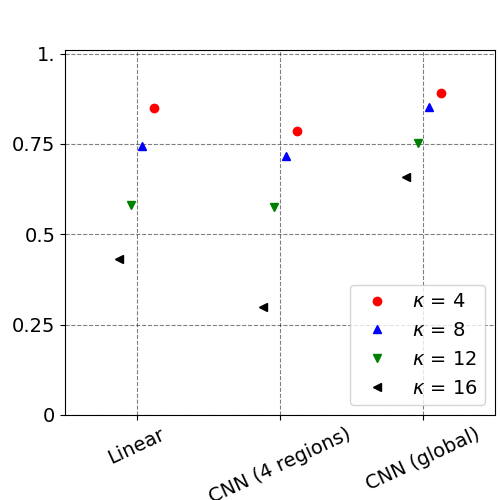}&
 \includegraphics[width=1\linewidth]{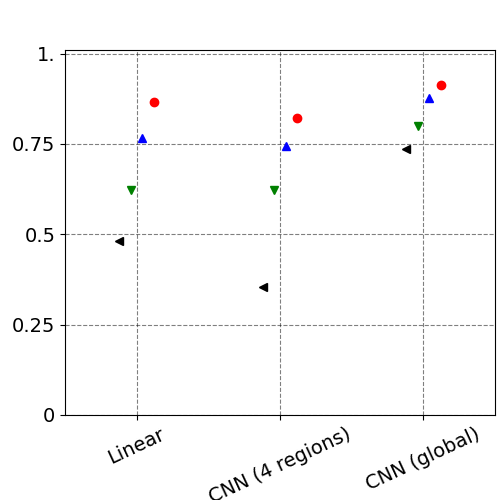}&
 \includegraphics[width=1\linewidth]{images/basic/r2_co2_0p01_filtering_gaussian_lossfun_heteroscedastic_training_filtering_gaussian_Stemp_r2.png}
\end{tabularx}
\caption{Test $R^2$ coefficient for longitudinal momentum (left), latitudinal momentum (center) and temperature (right) of the linear-inversion parameterization described in Section~\ref{sec:lsrp} and the CNN parameterization described in Section~\ref{sec:data-driven} trained on 4 regions and on the whole planet (global). Results for four different resolutions, corresponding to coarse-graining factors of 4, 8, 12 and 16 are shown. The CNN models were trained by minimizing the MSE loss (rows 1 and 2) and the Gaussian heteroscedastic loss (rows 3 and 4). Row 2 and 4 show results for the test set with increased $\text{CO}_2$ level. 
The data were processed using Gaussian filtering}
    \label{fig:r2-values-gauss}
\end{figure}

\begin{figure}[t]
\begin{tabularx}{0.1\textwidth} { 
>{\centering\arraybackslash}m{0.0001\linewidth} 
>{\centering\arraybackslash}m{0.22\linewidth}
>{\centering\arraybackslash}m{0.22\linewidth}
>{\centering\arraybackslash}m{0.22\linewidth}
>{\centering\arraybackslash}m{0.22\linewidth}}
&  
$\quad\quad\cgfac = 4$ &
$\quad\quad\cgfac = 8$ &
$\quad\quad\cgfac = 12$ & 
$\quad\cgfac = 16$\\
   \begin{turn}{90}$S_u$ density \end{turn} &
 \includegraphics[width=1\linewidth]{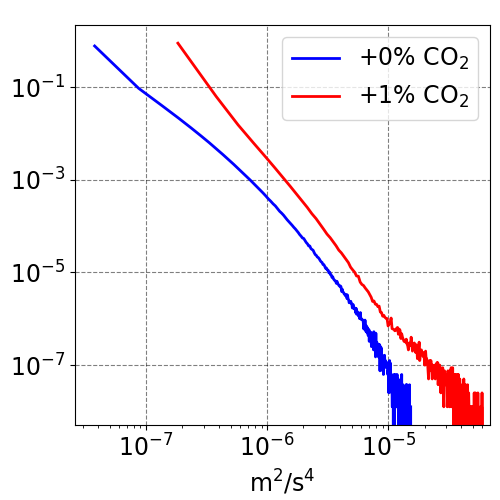} & 
 \includegraphics[width=1\linewidth]{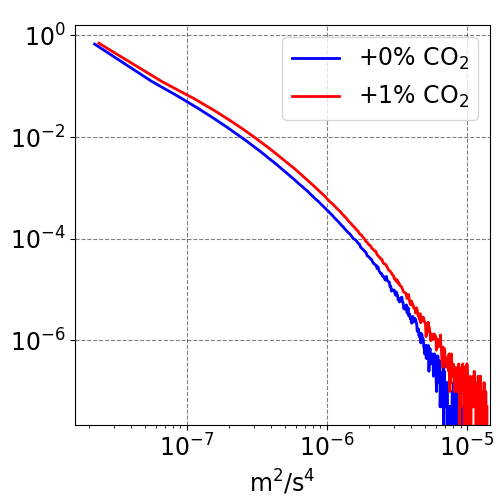} &
 \includegraphics[width=1\linewidth]{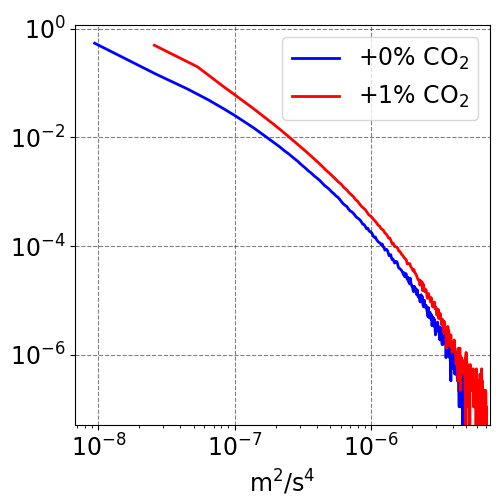}&
  \includegraphics[width=1\linewidth]{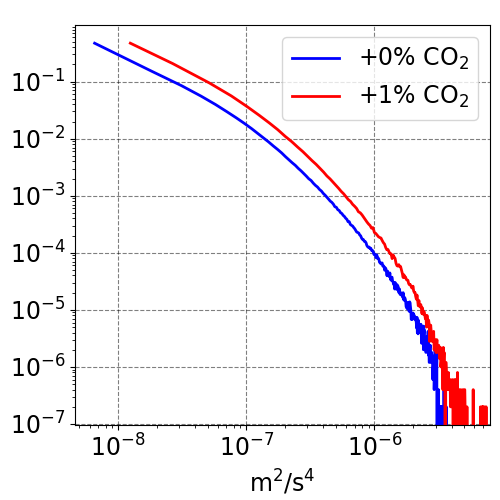}\\
 \begin{turn}{90}$S_v$ density \end{turn} &
  \includegraphics[width=1\linewidth]{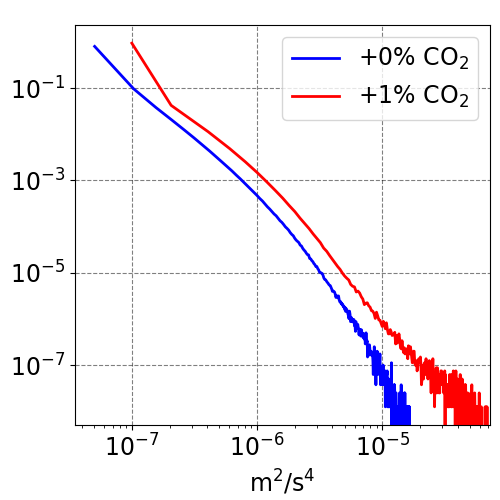} & 
 \includegraphics[width=1\linewidth]{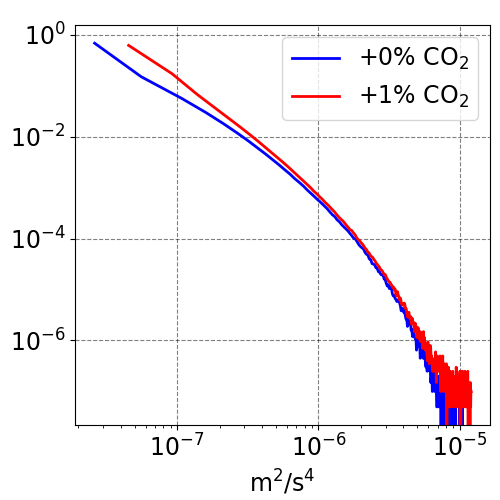} &
 \includegraphics[width=1\linewidth]{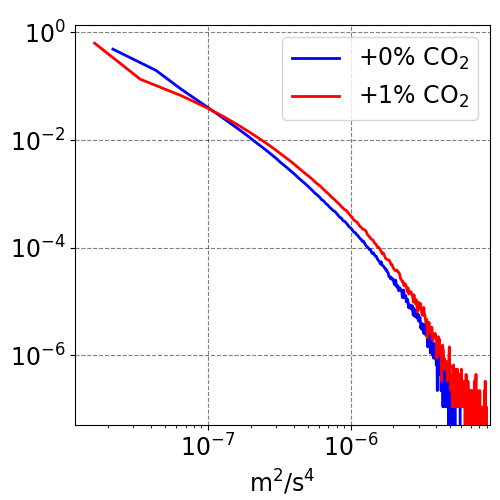}&
  \includegraphics[width=1\linewidth]{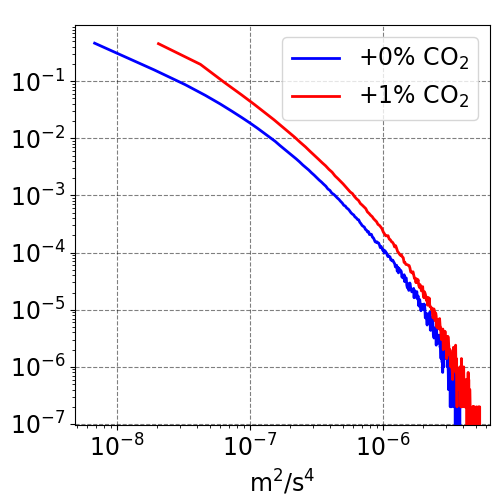}\\
   \begin{turn}{90}$S_T$ density \end{turn} &
  \includegraphics[width=1\linewidth]{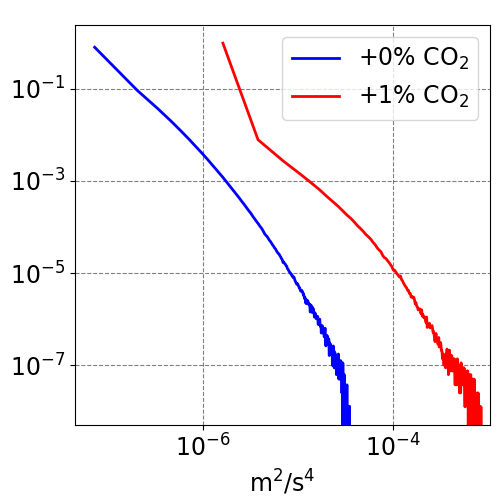} & 
 \includegraphics[width=1\linewidth]{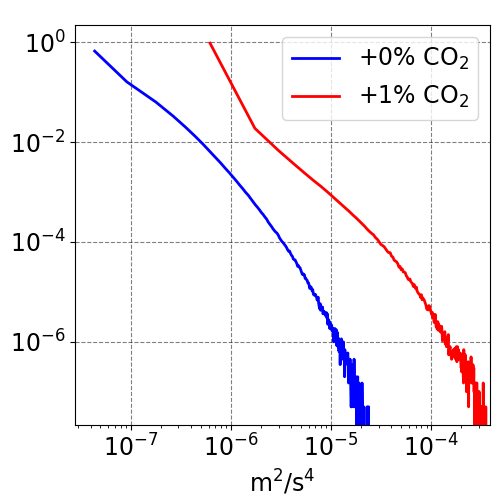} &
 \includegraphics[width=1\linewidth]{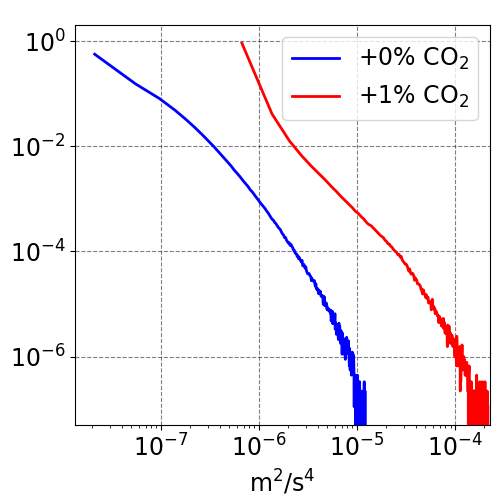}&
  \includegraphics[width=1\linewidth]{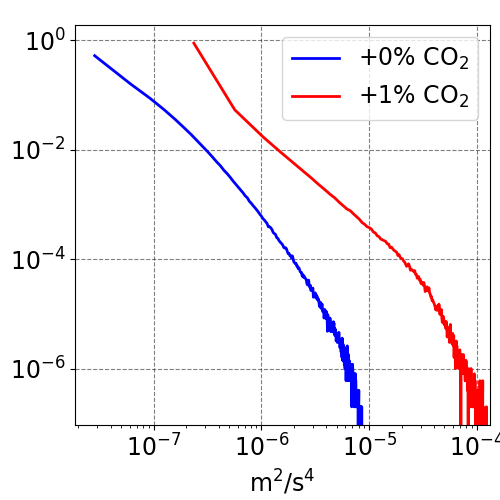}
\end{tabularx}
\caption{
    Distribution of the subgrid forcing for longitudinal momentum (top), latitudinal momentum (center) and temperature (bottom) for different coarse-graining factors (indicated by the value of $\cgfac$ above each column), computed from data at pre-industrial \cotwo{} levels (blue) and at significantly increased \cotwo{} levels (red). The data were processed using General-Circulation-Model filtering. We observe a clear distribution shift for the temperature data}
    \label{fig:forcing-distributions}
\end{figure}

Figure~\ref{fig:four-regions} shows the geographic locations used to train the 4-regions CNN model, following \mycite{guillaumin2021stochastic}. Figure~\ref{fig:coordinatewise-r2-global-all-sigmas} shows maps of the test $R^2$ coefficient for temperature of the CNN parameterization trained on the whole globe by minimizing the heteroscedastic Gaussian loss (left) and MSE (right). Figures~\ref{fig:r2-values-gcm} and \ref{fig:r2-values-gauss} show analogous results to Figure~\ref{fig:r_squared_MSE}, but for the heteroscedastic Gaussian loss and for Gaussian filtering, respectively. 
Figure~\ref{fig:forcing-distributions} shows the distribution shift in the subgrid forcing corresponding to momentum and temperature, between the control simulation and the simulation with increased \cotwo{} levels. 

\end{appendix}

\end{document}